\documentclass[a4paper,11pt,twoside]{article}
\usepackage{amsmath,amssymb,amsthm}
\usepackage[ps,dvips,frame,arc,knot]{xy}
\usepackage[dvips]{graphics}

\newcommand{\defeq}{:=}

\newcommand{\Z}{\mathbb{Z}}

\newcommand{\R}{\mathbb{R}}
\newcommand{\C}{\mathbb{C}}
\newcommand{\tens}{\otimes}  

\newcommand{\xd}{\mathrm{d}}  

\newcommand{\falg}{\mathcal{C}}  
\DeclareMathOperator{\id}{id}
\DeclareMathOperator{\tr}{tr}
\DeclareMathOperator{\Mor}{Mor}

\newcommand{\one}{\mathbf{1}}

\DeclareMathOperator{\cou}{\epsilon}
\DeclareMathOperator{\cop}{\Delta}
\DeclareMathOperator{\antip}{\mathrm{S}}
\DeclareMathOperator{\ev}{\mathrm{ev}}
\DeclareMathOperator{\coev}{\mathrm{coev}}
\DeclareMathOperator{\evi}{\widetilde{\mathrm{ev}}}
\DeclareMathOperator{\coevi}{\widetilde{\mathrm{coev}}}
\newcommand{\act}{\triangleright}

\renewcommand{\i}[1]{{}_{\scriptscriptstyle(#1)}}
\newcommand{\iu}[1]{{}_{\scriptscriptstyle(\underline #1)}}

\newcommand{\cR}{\mathcal{R}}
\newcommand{\cH}{\mathcal{H}}
\newcommand{\catmod}{{\mathcal{C}}}
\newcommand{\rep}[1]{\mathcal{R}(#1)}
\newcommand{\comod}[1]{\mathcal{M}^{#1}}
\DeclareMathOperator{\cdim}{loop}
\DeclareMathOperator{\sdim}{sdim}
\DeclareMathOperator{\qdim}{qdim}
\newcommand{\pf}{\mathcal{Z}}
\newcommand{\pfm}{\tilde{\mathcal{Z}}}

\theoremstyle{plain}
\newtheorem{prop}{Proposition}[section]
\newtheorem{lem}[prop]{Lemma}
\newtheorem{thm}[prop]{Theorem}

\newtheorem{dfn}[prop]{Definition}
\theoremstyle{definition}

\newcommand{\rxy}[1]{{\begin{xy} 0;<1mm,0mm>:<0mm,1mm>::0;0,#1
\end{xy}}}

\numberwithin{equation}{section}
\numberwithin{figure}{section}
\numberwithin{table}{section}

\allowdisplaybreaks

\begin{document}

\begin{titlepage}
\title{\textbf{Generalized Lattice Gauge Theory,\\ Spin Foams and
State Sum Invariants}}
\author{Robert Oeckl\footnote{email: oeckl@cpt.univ-mrs.fr}\\ \\
Centre de Physique Th\'eorique,
CNRS Luminy, Case 907,\\
F-13288 Marseille -- Cedex 9}
\date{CPT-2001/P.4258\\
29 October 2001}

\maketitle

\vspace{\stretch{1}}

\begin{abstract}
We construct a generalization of pure lattice gauge theory
(LGT) where the role of the gauge group is played by a tensor
category. The type of tensor category
admissible (spherical, ribbon, symmetric)
depends on the dimension of the underlying manifold ($\le 3$, $\le 4$, any).
Ordinary LGT is recovered if the category is the (symmetric) category of
representations of a compact Lie group.
In the weak coupling limit we recover discretized BF-theory in terms
of a coordinate free version of the spin foam formulation.
We work on general cellular
decompositions of the underlying manifold.

In particular, we are able to formulate LGT as well as spin foam
models of BF-type
with quantum gauge group (in dimension $\le 4$) and with supersymmetric
gauge group (in any dimension).

Technically, we express the partition function as a sum over diagrams
denoting morphisms in the underlying category.
On the LGT side this enables us to introduce a generalized
notion of gauge fixing corresponding to a topological move between
cellular decompositions of the underlying manifold.
On the BF-theory side this allows a rather geometric understanding
of the state sum invariants of Turaev/Viro, Barrett/Westbury and
Crane/Yetter which we recover.

The construction is extended to include Wilson loop and
spin network type observables as well as manifolds with
boundaries.
In the topological (weak coupling) case this leads to
TQFTs with or without embedded spin networks.

\end{abstract}

\vspace{\stretch{1}}
\end{titlepage}
\newpage
\tableofcontents
\newpage
\section{Introduction}

We start by describing the main motivations of the present work.

Lattice gauge theory (LGT) is our most successful approach to date at
describing the non-perturbative regime of the Standard Model, such as
bound states of QCD.
If the gauge group is abelian there is a well known duality
transformation exchanging the strong with the weak
coupling regime \cite{Sav:duality}. At the same time group valued
degrees of freedom are
replaced with character valued degrees of freedom.
As the latter also form a group
the dual theory is again a gauge theory living on the dual
lattice.
Even in the non-abelian case a ``dual'' formulation is
possible where the degrees of freedom are ``representation
valued''. While this is not a gauge theory anymore it is
known to be expressible as a (modified) spin foam model
\cite{Rei:worldsheet}. This dual model was explicitly constructed for
hypercubic lattices in \cite{OePf:dualgauge}. There, it was also shown
to be strong-weak dual to the ordinary formulation of LGT.
Thus, a better understanding of this formulation and improved
techniques to handle it appear of great value in order to extract a
strong-coupling expansion.

Quantum groups have their origin as symmetries of integrable
models. Thus, it is natural to ask whether they can be ``gauged'',
i.e., whether one could formulate gauge theories with quantum
groups. Indeed, this is supported e.g.\ by an analysis of Chern-Simons
theory where a necessary regularization of the path integral naturally
leads to quantum gauge groups \cite{Wit:qftjones}.
At the non-perturbative level lattice gauge theory appears
clearly the most suitable starting point for such a development.
Indeed, a proposal for a $q$-deformed LGT in 3 dimensions has been made
\cite{Bou:qlgt}. In 4 dimensions a generalized LGT for ribbon
categories was constructed on simplicial decompositions of the
underlying manifold \cite{Pfe:4dlgtrib}. A unified approach,
preferably for general cellular decompositions, is clearly desirable.

Spin foams have emerged as a description of space-time both in the
canonical loop approach to quantum gravity as well as in covariant
path integral approaches \cite{Bae:spinfoams}.
On the other hand, pure quantum gravity in 3 dimensions
turns out to be essentially quantum BF-theory, which is topological
\cite{Wit:gravsolv}. Indeed, a well defined path integral description
of BF-theory can be given in the spin foam framework.
A predecessor to these ideas is the state sum model of Ponzano and Regge
\cite{PoRe:limracah}.
$q$-deformations of the gauge group come about when a cosmological
constant is included \cite{MaSm:qdefqgrav}.
Spin foam models of BF-theory have recently been the starting point
for several proposals for quantum gravity also in 4 dimensions
\cite{BaCr:relsnet}.
See also the review \cite{Bae:introsfoam}.

A completely new type of algebraic topology
started to emerge in the 80s, initiated by the application of quantum
field theoretic ideas to low dimensional topology.
In particular, this led to new kinds of
topological invariants of
manifolds and the notion of topological quantum field theory (TQFT).
The most prominent invariants are the surgery invariant of Reshetikhin
and Turaev in 3 dimensions \cite{ReTu:inv3qg}, the state sum invariant
of Turaev and Viro in 3 dimensions \cite{TuVi:inv3,BaWe:invplm}
and the state sum invariant of
Crane and Yetter in 4 dimensions \cite{CrYe:4dtqft,CrKaYe:inv4}.
All those invariants require a (quantum) group or, more generally, a
certain type of category as input. It turns out that using ordinary
groups (as compared to quantum groups) does not lead to interesting
new invariants. The reason for this can be seen to lie in the fact
that the quantum groups ``feel'' more about the topology than the
ordinary groups. However, this remains somewhat obscure in the
standard approaches to the invariants.

This work aims to contribute to the above-mentioned developments as
well as to improve our understanding of the connections
between them.

We construct a generalization of pure lattice gauge theory where the
role of the gauge group is played by a monoidal (or ``tensor'') category.
Ordinary LGT is recovered if the category is taken to be the category of
representations of a compact Lie group.
We make heavy use of the relation between (types of) tangle
diagrams and (types of) monoidal categories as developed in
\cite{FrYe:braidcat,ReTu:ribboninv,BaWe:sphcat}.
We suitably extend this to a diagrammatic calculus which allows to
express the partition function of LGT in a purely diagrammatic way. More
concretely, the diagram defining the partition function as a morphism
in the given category is constructed
from a cellular decomposition of the underlying manifold. This diagram
can be considered as the ``projection'' onto the plane of a graph
embedded into the manifold.

The cases of symmetric and nonsymmetric categories are different in an
essential way. In the former case a lattice (combinatorial 2-complex)
is sufficient to define LGT. In particular, this lattice can be
obtained as the dual 2-skeleton of a cellular decomposition of a
manifold of arbitrary dimension which need not be orientable. In the
nonsymmetric case an orientable manifold is required and the
orientation indeed enters into the construction of LGT. Furthermore, the
dimension of the manifold is restricted by the type of
category. Concretely, the maximal allowed dimension is 2, 3, 4 for
pivotal, spherical, ribbon categories respectively. Indeed, the
geometric nature of our construction makes this connection
between the dimension and the admissible type of category
rather transparent through the isotopy properties characterizing
the tangle diagrams associated with the category.

The gauge invariance properties of conventional LGT
are shown to extend to our generalized LGT. Gauge fixing can be
reexpressed as the invariance of the partition function under a
certain topological move relating different cellular decompositions of
the manifold.
The standard Wilson loop and spin network observables are included in
our generalized LGT. However, in the nonsymmetric case the maximal
allowed dimension for a given type of category drops by one.
Extending our formulation to manifolds with boundaries we obtain (as
expected) spin
networks as states on the boundary and consider briefly the
construction of the associated TQFTs.

In the weak coupling limit we obtain (discretized) BF-theory and
recover the abovementioned state sum invariants of Turaev and Viro
\cite{TuVi:inv3}, Barrett and Westbury \cite{BaWe:invplm} and Crane
and Yetter \cite{CrYe:4dtqft,CrKaYe:inv4}.

Section~\ref{sec:catdiag} introduces the relevant types of categories, their
diagrammatics and the notion of semisimplicity.
Section~\ref{sec:reptheo} reviews how those categories arise as
categories of representations of groups, supergroups and quantum
groups. Furthermore, the
diagrammatic calculus as well as the notion of semisimplicity is
further developed for these cases.
The partition function of generalized LGT is introduced in
Section~\ref{sec:lgt}. First, a diagrammatic representation of the
partition function of ordinary LGT is derived. Then, the
generalization to different types of categories is performed.
Gauge symmetry and gauge fixing are considered in
Section~\ref{sec:gauge}. Observables of Wilson loop and spin network
type are implemented in Section~\ref{sec:obs}. The partition function
is extended to manifolds with boundaries in
Section~\ref{sec:boundary}. (Generalized) spin networks emerge as
boundary states and the construction of the relevant TQFTs in the
topological case is sketched.
Special cases are considered in Section~\ref{sec:specas}. In
particular, we consider how the spin foam picture emerges. We discuss the
topological weak coupling limit (BF-theory) and the specialization
to the various state sum invariants.
An outlook is presented in Section~\ref{sec:outlook}.

Propositions and lemmas for which the proof is not included are either
to be found in the given references or verified straightforwardly.

\section{Categories and Diagrams}
\label{sec:catdiag}

In this section we introduce the various types of categories which
are to play the role of the ``gauge symmetry'' for generalized
LGT. Furthermore, we introduce the associated diagrammatics which is
instrumental in our formulation of LGT.

\subsection{Monoidal Categories with Structure}

We start by introducing the necessary categorial notions. A standard
reference for general category theory and monoidal categories is
\cite{Mac:categories}.
Pivotal, spherical and ribbon categories as well as their
diagrammatics are introduced in
\cite{FrYe:braidcat,BaWe:sphcat,ReTu:ribboninv}.

In the following, category is always taken to mean $\C$-linear
category. By this we mean that the set $\Mor(V,W)$ of morphisms from
an object $V$ to an object $W$ forms a vector space over the field
$\C$. Furthermore, the composition of morphisms
\[
 \Mor(U,V)\times\Mor(V,W)\to\Mor(U,W),\quad (f, g) \mapsto g\circ f
\]
is required to be bilinear.

\begin{dfn}
A (strict) \emph{monoidal category} is a category $\catmod$ together
with a bifunctor $\tens:\catmod\times\catmod\to\catmod$ (called
\emph{tensor product}) and a choice of
unit element $\one\in\catmod$.
Furthermore we require the equalities 
$(U\tens V)\tens W = U\tens (V\tens W)$ and $U\tens\one = U =
\one\tens U$. We also require that $\Mor(\one,\one)=\C$
and the monoidal structure on morphisms be identified with their
tensor product as vectors.
\end{dfn}
\begin{dfn}
A \emph{rigid monoidal category} is a monoidal category $\catmod$
together with a contravariant functor $*:\catmod\to\catmod$ called
\emph{dual} and
morphisms $\ev_V:V^*\tens V\to\one$ \emph{(evaluation)},
$\coev_V:\one\to V\tens V^*$ \emph{(coevaluation)} such
that
\begin{gather*}
(\id_V\tens\ev_V)\circ (\coev_V\tens\id_V)=\id_V,\quad
(\ev_V\tens\id_{V^*})\circ (\id_{V^*}\tens\coev_V) =\id_{V^*},
\end{gather*}
and for a morphism $\Phi:V\to W$ we have its dual $\Phi^*:W^*\to V^*$
given by
\[
 \Phi^*=(\ev_W\tens \id_{V^*})\circ
 (\id_{W^*}\tens\Phi\tens\id_{V^*})\circ (\id_{W^*}\tens\coev_V) .
\]
\end{dfn}
\begin{dfn}
\label{def:pivcat}
Let $\catmod$ be a rigid monoidal category
together with a natural equivalence $\tau_V:V\mapsto V^{**}$ 
such that $\tau_V\tens\tau_W=\tau_{V\tens W}$ and
$\tau_{V^*}^{-1}=(\tau_V)^*$.
Define $\evi_V:V\tens V^*\to\one$ and
$\coevi_V:\one\to V^*\tens V$ as
\[
 \evi_V\defeq \ev_{V^*}\circ (\tau_V\tens\id_{V^*}),\quad
 \coevi_V\defeq (\id_{V^*}\tens\tau_V^{-1})\circ\coev_{V^*} .
\]
If for all morphisms $\Phi:V\to W$ the equality
\[
  \Phi^*=(\id_{V^*}\tens\evi_W)\circ
 (\id_{V^*}\tens\Phi\tens\id_{W^*})\circ (\coevi_V\tens\id_{W^*})
\]
holds we call $\catmod$ a \emph{pivotal category}.
For a morphism $\Phi:V\to V$ define $\tr_-(\Phi):\one\to\one$ and
$\tr_+(\Phi):\one\to\one$ as
\[
 \tr_-(\Phi)\defeq\ev_V\circ (\id_{V^*}\tens\Phi) \circ\coevi_V,\quad
 \tr_+(\Phi)\defeq\evi_V\circ (\Phi\tens\id_{V^*}) \circ\coev_V .
\]
For an object $V$ define morphisms $\one\to\one$ as
\[
 \cdim_- V\defeq \tr_-(\id_V), \quad \cdim_+ V\defeq \tr_+(\id_V).
\]
\end{dfn}

\begin{lem}
\label{lem:pivtr}
In a pivotal category the identities $\tr_-(\Phi)=\tr_+(\Phi^*)$ and
$\tr_+(\Phi)=\tr_-(\Phi^*)$ hold for any morphism $\Phi$.
\end{lem}

\begin{dfn}
\label{def:sphcat}
A \emph{spherical category} is a pivotal category such that for all
objects $V$ and morphisms
$\Phi:V\to V$ the equality $\tr_-(\Phi)=\tr_+(\Phi)$ holds.
Write $\tr\defeq\tr_+=\tr_-$ and $\cdim\defeq\cdim_+=\cdim_-$.
\end{dfn}

\begin{dfn}
A \emph{braided monoidal category} is a monoidal category together
with a natural equivalence $\psi_{V,W}:V\tens W\to W\tens V$ (called
\emph{braiding}) such that
the following conditions hold:
\begin{gather*}
\psi_{U\tens V, W}=(\psi_{U,W}\tens\id_V)\circ(\id_U\tens\psi_{V,W}),\\
\psi_{U, V\tens W}=(\id_V\tens\psi_{U,W})\circ(\psi_{U,V}\tens\id_W),\\
\psi_{V,\one}=\id_V=\psi_{\one,V} .
\end{gather*}
\end{dfn}
\begin{dfn}
A \emph{ribbon category} is a rigid braided monoidal category together with
a natural equivalence $\nu_V:V\to V$ such that
\[
\nu_{V\tens W}=\psi^{-1}_{V,W}\circ\psi^{-1}_{W,V} (\nu_V\tens\nu_W),
\quad \nu_\one=\id_\one,\quad \nu_{V^*}=(\nu_V)^*.
\]
\end{dfn}
\begin{lem}
A ribbon category is a spherical category by setting
\[
 \tau_V\defeq (\ev_V\tens\id_{V^{**}})\circ
 (\psi^{-1}_{V^*,V}\tens\id_{V^{**}})
 \circ (\nu_V\tens\coev_{V^*}) .
\]
\end{lem}
\begin{dfn}
\label{def:symcat}
A \emph{symmetric category} is a rigid monoidal category together
with a natural equivalence $\psi_{V,W}:V\tens W\to W\tens V$ such that
the following conditions hold:
\begin{gather*}
\psi_{U\tens V, W}=(\psi_{U,W}\tens\id_V)\circ(\id_U\tens\psi_{V,W}),\\
\psi_{U, V\tens
W}=(\id_V\tens\psi_{U,W})\circ(\psi_{U,V}\tens\id_W),\\
\psi_{W,V}\circ\psi_{V,W}=\id_{V\tens W}, \qquad
\psi_{V,\one}=\id_V=\psi_{\one,V} .
\end{gather*}
\end{dfn}
Note that the usual definition of ``symmetric'' does not imply
rigidity. We include it here for uniformity of terminology.
\begin{lem}
A symmetric category is a ribbon category by noting that the symmetric
structure $\psi$ is a self-inverse braiding and by setting
$\nu_V\defeq\id_V$.
\end{lem}

\subsection{Diagrams and Isotopy Invariance}
\label{sec:diag}

The types of categories we will be mainly interested in are the
\emph{pivotal}, \emph{spherical}, \emph{ribbon}, and \emph{symmetric}
categories. Morphisms in those categories can be conveniently
described by directed tangle diagrams with additional structure.
Remarkably, the denoted morphisms remain invariant under certain
isotopies of these diagrams. This plays a key role in the construction
of the partition function of generalized LGT. We introduce this
diagrammatics in the present section.

We start by considering the pivotal case.
A diagram (without coupons) consists of a finite number of
non-intersecting lines in the plane.
The lines end at the top or bottom line of the diagram or form closed
loops. Each line carries an object label and an arrow. A
diagram as a whole defines a morphism in the category. If the
object labels of the lines
ending at the top are $V_1,\dots,V_n$ and the ones ending at the
bottom are $W_1,\dots,W_m$ it defines a morphism
$V_1\tens\dots\tens V_n\to W_1\tens\dots\tens W_m$. For lines with
an arrow pointing upward the respective object is replaced by
its dual.

\begin{figure}
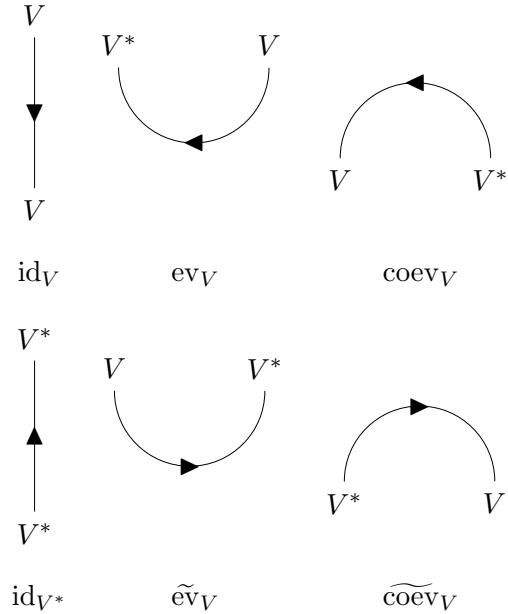

\begin{center}
\begin{tabular}{cccc}
\input{figures/fig_tan_id}
& \input{figures/fig_tan_archdl} & \input{figures/fig_tan_archul} \\
\\
$\id_V$ &  $\ev_V$ & $\coev_V$ \\
\\
\input{figures/fig_tan_ids} & \input{figures/fig_tan_archdr} &
\input{figures/fig_tan_archur} \\
\\
$\id_{V^*}$ & $\evi_V$ & $\coevi_V$
\end{tabular}
\caption{Elementary tangle diagrams and their assigned morphisms.}
\label{fig:elemtan}
\end{center}
\end{figure}

For elementary diagrams the assignments are listed in
Figure~\ref{fig:elemtan}. Note that the unit object $\one$ is usually
not explicitly represented in the diagrams.
Diagrams placed horizontally next to each
other correspond to the tensor product of morphisms. For more
complicated diagrams the morphism is obtained by slicing the
diagram horizontally into elementary slices and composing the
corresponding morphisms from top to bottom.

We need to introduce another elementary diagram: a
\emph{coupon}. This is a rectangle which is connected to a certain
number of lines on the top and on the bottom, see
Figure~\ref{fig:tcoupon}. Furthermore, it carries a label denoting a
morphism from the tensor product of the objects (or
their duals) labelling the lines at the top to the corresponding
tensor product at the bottom. Under the assignment of
morphisms to diagrams a coupon is simply assigned the morphism
with which it is labeled.

\begin{figure}
\begin{center}
\input{figures/fig_tan_coupon}
\caption{Coupon.}
\label{fig:tcoupon}
\end{center}
\end{figure}

The morphism associated with such a diagram is invariant under planar
isotopy. That is, any diagram that is related to a given one by an
isotopy in $\R^2$ (holding the endpoints of lines at the top and
bottom fixed) yields the same morphism. Note that a closed diagram,
i.e., a diagram without endpoints denotes a morphism $\one\to\one$ and
thus an element in $\C$.

\begin{figure}
\begin{center}
\input{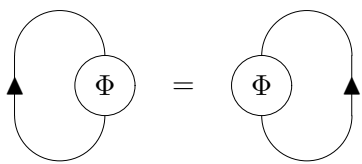}
\caption{Trace property defining a spherical category.}
\label{fig:sphprop}
\end{center}
\end{figure}

We now turn to the spherical case. The additional property of a
spherical category as compared to a pivotal one
(Definition~\ref{def:sphcat}) can be easily expressed diagrammatically
(Figure~\ref{fig:sphprop}). The consequence is an enhanced isotopy
invariance of the diagrammatics. That is, given a closed diagram
inscribed on a 2-sphere, any isotopic deformation followed by piercing
the 2-sphere at some point to identify it with the plane yields the same
morphism.

For ribbon categories we need to modify the diagrammatics as
follows. Instead of lines we now consider ribbons. One can think of
this as equipping the lines with a \emph{framing}.
The orientation of
a ribbon (i.e. the framing) at its endpoints is always ``face up''.
In particular, a ribbon has an ``upside'' and a ``downside'' and thus an
orientation. This is also true for ribbon loops. E.g.,
a M\"obius strip is not allowed.

\begin{figure}
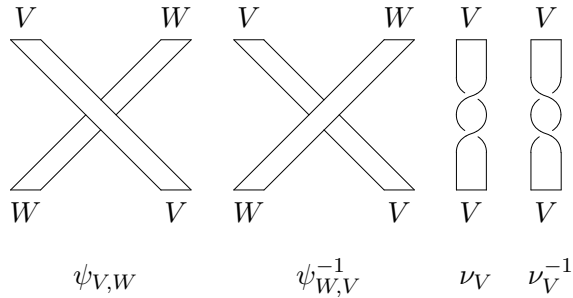

\begin{center}
\begin{tabular}{cccc}
\input{figures/fig_rib_xo} & \input{figures/fig_rib_xu}
& \input{figures/fig_rib_tpos} & \input{figures/fig_rib_tneg} \\
\\
$\psi_{V,W}$ & $\psi^{-1}_{W,V}$
& $\nu_V$ & $\nu_V^{-1}$
\end{tabular}
\caption{Additional ribbon diagrams and their assigned morphisms.}
\label{fig:elemrib}
\end{center}
\end{figure}

We also have additional elementary diagrams in the ribbon case. One
must certainly be a twist of the ribbon (and its inverse). Furthermore,
we allow crossings of ribbons, with a distinction between over- and
under-crossings. See Figure~\ref{fig:elemrib}
(where the arrows are omitted in the diagrams).

The isotopy invariance of the ribbon diagrammatics is even stronger
than in the spherical case. Indeed, we can think of a diagram as
obtained by projecting a ribbon tangle in $\R^3$ (or $S^3$) onto the
plane. Then, the projection of any isotopic ribbon tangle in $\R^3$ (or
$S^3$) will yield a diagram corresponding to the same morphism.
This is because (the ribbon version of) the Reidemeister moves give
rise to identities of morphisms.
Note that in case of an open diagram, i.e., a diagram with endpoints,
the endpoints are to be held fixed and
no isotopy involving the moving of a ribbon ``around''
end points is allowed.

When drawing ribbon diagrams it is sometimes convenient and sufficient
to just draw lines instead of ribbons. The convention in this case is
that a line represents a ribbon which lies everywhere ``face
up''. This is called the \emph{blackboard framing}.

Note that as a ribbon category is in particular a spherical category,
we can convert a diagram for a morphism in the latter into a diagram
for the same morphism in the former. This is simply achieved by
introducing the blackboard framing.

\begin{figure}
\begin{center}
\input{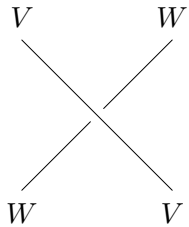}
\caption{The crossing $\psi_{V,W}$ in the symmetric case.}
\label{fig:symbraid}
\end{center}
\end{figure}

Finally we consider the case of symmetric categories. The
diagrammatics is again similar to the pivotal and spherical
cases. That is, we have again lines instead of ribbons. The only
difference is that crossings are allowed, and there is only one
type of crossing. See Figure~\ref{fig:symbraid}.

The invariance properties of the diagrammatics are the strongest in
the symmetric case. Consider a set of coupons, a set of end points at the
top and bottom and a specification of which end point (on the border
of the diagram or on a coupon) is to be connected to which other one
and with which arrow direction. This combinatorial data
already specifies a morphism in the category. That is, any diagram that
satisfies this combinatorial data yields the same morphism.

A symmetric category is in particular a ribbon category.
Thus, we can convert the diagram for a morphism in the latter into the
diagram for the same morphism in the former. This is simply achieved
by removing the framing (as the twist is now trivial) and
forgetting about the distinction
between over- and under-crossings which become identical.

\begin{table}
\begin{center}
\begin{tabular}{|l|l|}
\hline
type of category & diagrammatic invariance\\
\hline
pivotal & isotopy in $\R^2$\\
spherical & isotopy in $S^2$\\
ribbon & isotopy in $\R^3$, $S^3$\\
symmetric & combinatorial\\
\hline
\end{tabular}
\caption{Diagrammatic invariance for different types of categories.}
\label{tab:cat}
\end{center}
\end{table}

The invariance properties of the diagrammatics for the different types
of categories are summarized in Table~\ref{tab:cat}.
By slight abuse of terminology we refer to graphs which do
not live in the plane but are embedded into a manifold or lattice as
diagrams in the same way (although they do not directly define a
morphism). If an explicit distinction is necessary we
refer to these as \emph{embedded diagrams} and to planar ones that are
obtained from these as \emph{projected diagrams}.

Although having historically a more restricted meaning, we define the
term \emph{spin network} here to mean precisely a diagram as
considered above. Thus, there are different types of spin networks
depending on the type of category. The original version is for the
representation category of the group $SU(2)$
\cite{Pen:combst}. Furthermore, there one considers
only lines labeled by irreducible representations and one type of
coupon (represented by a trivalent vertex), which is a suitably
normalized intertwiner between three incident representations.

\subsection{Semisimplicity}

Recall that an object $V$ is called simple if $\Mor(V,V)\cong\C$
as a vector space. Usually one defines a
category to be semisimple if any object decomposes into a direct sum
of simple objects. However,
we need to adopt a more general definition which does not require
direct sums \cite{Tur:qinv}.

\begin{dfn}
\label{def:semisimple}
A category is called \emph{semisimple} if for each object $V$ there
exists a finite set of simple objects $V_i$ and morphisms
$f_i:V\to V_i$, $g_i:V_i\to V$ such that
\[
\id_V=\sum_i g_i\circ f_i .
\]
We call this
data also a \emph{decomposition} of $V$.
\end{dfn}

\begin{prop}
\label{prop:proj}
Let $\catmod$ be a semisimple category. We define a
morphism $T_V:V\to V$ for each object $V$ as follows. Let $V_i$,
$f_i:V\to V_i$, $g_i:V_i\to V$ with $i\in I$ be a decomposition of
$V$. Let $I'\defeq \{i\in I | V_i\cong \one\}$. Then
\[
 T_V\defeq \sum_{i\in I'} g_i\circ f_i .
\]
This definition is well (independent of the decomposition) and gives
rise to the following properties:
\begin{itemize}
\item[(a)] $T_\one = \id_\one$.
\item[(b)] $T_V=0$ for $V$ simple and $V\ncong \one$.
\item[(c)] $T$ is a projector, i.e.\ $T_V^2=T_V$.
\item[(d)] $T$ defines a natural transformation of the identity
functor with itself. That is, for $\Phi:V\to W$ a morphism we have
$T_W\circ\Phi=\Phi\circ T_V$.
\end{itemize}
If furthermore $\catmod$ is monoidal:
\begin{itemize}
\item[(e)] $T_V\tens T_W=T_{V\tens W}\circ (T_V\tens\id_W)
 =T_{V\tens W}\circ (\id_V\tens T_W)$.
\end{itemize}
If $\catmod$ is rigid monoidal:
\begin{itemize}
\item[(f)] $T$ is self-dual, i.e.\ $(T_V)^*=T_{V^*}$.
\end{itemize}
If $\catmod$ is ribbon:
\begin{itemize}
\item[(g)] 
 $\psi_{V,W}\circ (T_V\tens\id_W) = \psi^{-1}_{W,V}\circ
 (T_V\tens\id_W)$.
\item[(h)] $\nu_V\circ T_V= T_V$.
\end{itemize}
\end{prop}
\begin{proof}
We start by showing that $T$ is well defined. Let
$\{V_i,f_i,g_i\}_{i\in I}$ and
$\{\tilde{V}_j,\tilde{f}_j,\tilde{g}_j\}_{j\in J}$ be two
decompositions of the object $V$. Define $I'\defeq \{i\in I\, |\,
V_i\cong\one\}$ and $J'\defeq \{j\in J\, |\,
\tilde{V}_j\cong\one\}$. We need to show that $T_V\defeq \sum_{i\in
I'} g_i\circ f_i$ and
$\tilde{T}_V\defeq \sum_{j\in J'} g_j\circ f_j$ are equal. Since $V_i$
and $\tilde{V}_j$ are simple objects any morphism $\tilde{f}_j\circ
g_i$ must be zero if $V_i\ncong \tilde{V}_j$. This implies
\[
 T_V=\sum_{i\in I', j\in J}
\tilde{g}_j\circ\tilde{f}_j\circ g_i\circ f_i
=\sum_{i\in I', j\in J'} \tilde{g}_j\circ\tilde{f}_j\circ g_i\circ f_i
=\sum_{i\in I, j\in J'} \tilde{g}_j\circ\tilde{f}_j\circ g_i\circ f_i
=\tilde{T}_V .
\]
Thus, $T_V$ is well defined. Note that the above expression also
proofs the projection property (c), as the term in the middle is
$\tilde{T}_V\circ T_V= T_V^2$. The properties (a) and (b) follow
immediately by taking the canonical decomposition of a simple object.
The proof of (d) is a small modification of the proof of well
definedness. We now have two different objects $V, W$ and a morphism
$\Phi:V\to W$ sandwiched in between. Considering decompositions of $V$
and $W$ we get as above $\Phi\circ T_V=T_W\circ\Phi\circ
T_V=T_W\circ\Phi$.

Now assume $\catmod$ to be monoidal. Let $\{V_i,f_i,g_i\}_{i\in I}$
and $\{W_j,p_j,q_j\}_{j\in J}$ be decompositions of the objects $V$
and $W$ and $\{U_k, a_k, b_k\}_{k\in K}$ a decomposition of
$V\tens W$. Consider the composition $b_k\circ a_k \circ (g_i\tens
q_j)\circ (f_i\tens p_j)$. Observe that $a_k \circ (g_i\tens q_j)$
vanishes if two of the objects $V_i, W_j, U_k$ are isomorphic to
$\one$ while the third one is not. Thus, defining the restricted index
sets as above, summing over $I', J', K$ or $I', J, K'$ or $I, J', K'$
yields the same morphism. This proofs (e).

Now assume $\catmod$ to be rigid. Let
$\{V_i,f_i,g_i\}_{i\in I}$ be a decomposition of the object $V$. Then
$\{V^*_i,g^*_i,f^*_i\}_{i\in I}$ is a decomposition of $V^*$. As
$V^*_i\cong\one$ iff $V_i\cong\one$, this implies
property (f).

Now assume $\catmod$ to be ribbon. Properties (g) and (h) follow
by the naturality of $\psi$ and $\nu$ and their properties
$\psi_{\one,W}=\id_W=\psi^{-1}_{W,\one}$ and $\nu_\one=\id_\one$.
\end{proof}

\begin{figure}
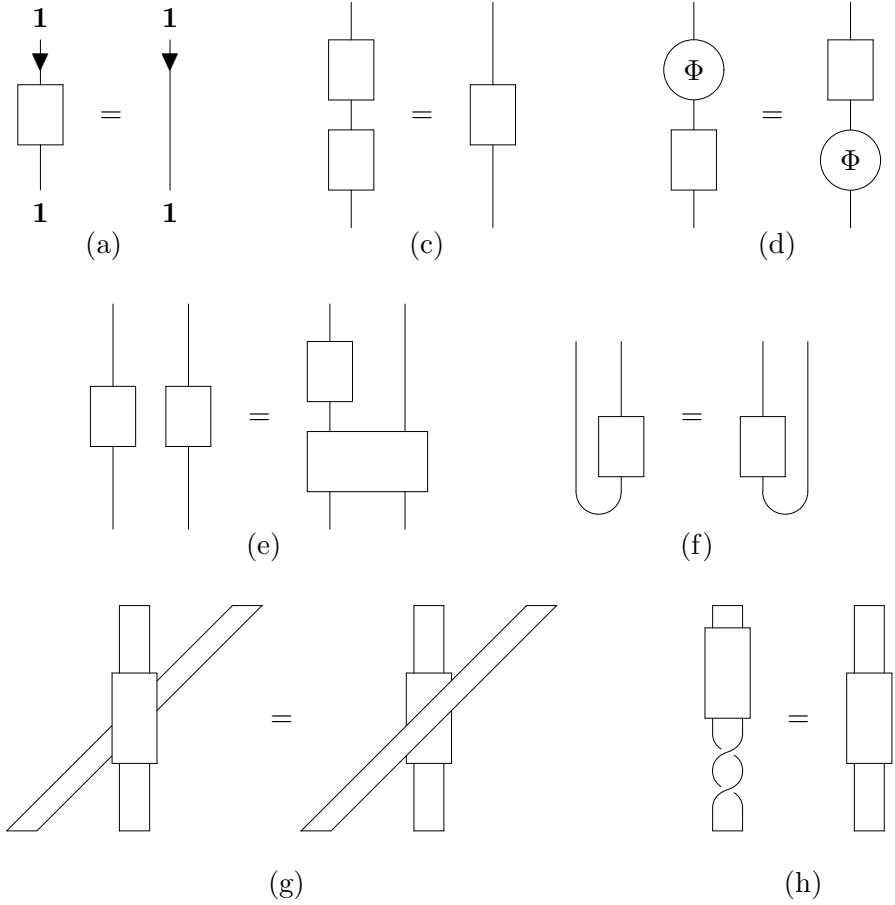

\begin{center}
\begin{tabular}{cp{1cm}cp{1cm}c}
\input{figures/fig_p_unit} && \input{figures/fig_p_proj} 
&& \input{figures/fig_p_com} \\
(a) && (c) && (d)
\end{tabular}
\\ \vspace{5mm}
\begin{tabular}{cp{1cm}c}
\input{figures/fig_p_comp} && \input{figures/fig_p_sd} \\
(e) && (f)
\end{tabular}
\\ \vspace{5mm}
\begin{tabular}{cp{1cm}c}
\input{figures/fig_qp_braid} && \input{figures/fig_qp_twist} \\
\\
(g) && (h)
\end{tabular}
\end{center}
\caption{Properties of $T$.}
\label{fig:propT}
\end{figure}

Using the diagrammatic language introduced above we can represent the
morphism $T$ by a coupon. As $T$ is defined for any object we
represent it simply by a coupon without label. The properties of $T$
can be expressed as diagrammatic identities, see
Figure~\ref{fig:propT}.

As this will be of importance later, we note that due to
``factorization'' of $T$ though the unit object $\one$ we can define a
``braiding'' composed with $T$ also in a general monoidal category.

\begin{dfn}
Let $\catmod$ be a semisimple monoidal category. Let $V$, $W$ be
objects in $\catmod$. Let $V_i$,
$f_i:V\to V_i$, $g_i:V_i\to V$ with $i\in I$ be a decomposition of
$V$, $I'\defeq \{i\in I | V_i\cong \one\}$. We define
$\psi_{T(V),W}:V\tens W\to W\tens V$ and $\psi_{W,T(V)}:W\tens V\to
V\tens W$ as follows.
\begin{gather*}
\psi_{T(V),W}\defeq\sum_{i\in I'} (\id_W\tens f_i)\circ (g_i\tens\id_W),\\
\psi_{W,T(V)}\defeq\sum_{i\in I'} (f_i\tens\id_W)\circ (\id_W\tens g_i).
\end{gather*}
\end{dfn}

\begin{figure}
\begin{center}
\input{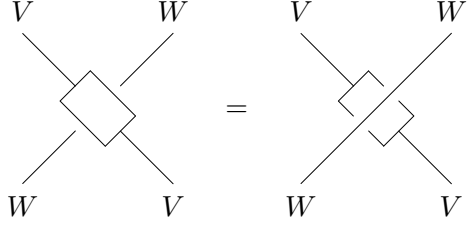}
\end{center}
\caption{The $T$-braiding $\psi_{T(V),W}$. The equality
indicates that there is just one type of crossing -- no distinction
between ``over'' and ``under''.}
\label{fig:Tcross}
\end{figure}

\begin{figure}
\begin{center}
\input{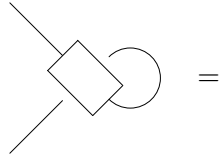}
\end{center}
\caption{``Twisting identity'' for the $T$-braiding.}
\label{fig:r1T}
\end{figure}

This definition simply uses the property $\one\tens
W=W\tens\one$.
Obviously, in the ribbon (or symmetric) case
$\psi_{T(V),W}=\psi_{V,W}\circ (T_V\tens\id_W)$ and
$\psi_{W,T(V)}=\psi_{W,V}\circ (\id_W\tens T_V)$. We can represent
this \emph{$T$-braiding} diagrammatically as in
Figure~\ref{fig:Tcross}. This can be considered as an additional
elementary diagram in the pivotal and spherical case.
Note that its properties are analogous to those of a
braiding in a symmetric category. In particular, we have the identity
depicted in Figure~\ref{fig:r1T}. (This follows by writing the
$T$-morphism diagrammatically as a decomposition and using invariance
under planar isotopy.)
Indeed, this can be considered the generalization of property (h) of
Proposition~\ref{prop:proj} to the non-ribbon case.

Important identities for the $T$-morphism on a tensor product are the
following.

\begin{figure}
\begin{center}
\input{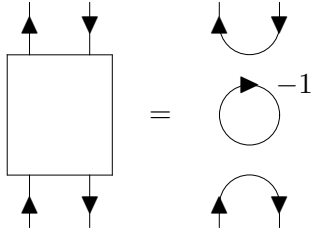}
\end{center}
\caption{Identity for the $T$-morphism on a tensor product of a simple
object with its dual. All lines are labeled by the same object.}
\label{fig:dualid}
\end{figure}

\begin{prop}
\label{prop:dualid}
Let $\catmod$ be a semisimple pivotal category.
For $V$ simple we have $\cdim_\pm V\neq 0$.
For two inequivalent simple objects $V$ and $W$ the morphisms $T_{V^*\tens
W}$ and $T_{W\tens V^*}$
are zero. Furthermore, for $V$ simple we have the identities
\[
T_{V^*\tens V}=\coevi_V\circ(\cdim_- V)^{-1}\circ\ev_V ,\quad
T_{V\tens V^*}=\coev_V\circ(\cdim_+ V)^{-1}\circ\evi_V .
\]
The first one is diagrammatically represented in
Figure~\ref{fig:dualid} while the second one is obtained by reversing
all arrows.
\end{prop}
\begin{proof}
Let $V$ and $W$ be simple objects. We can use $\coev_V$ to identify
the morphism spaces $\Mor(V^*\tens W,\one)$ and $\Mor(W, V)$. Thus, by
simplicity the dimension of $\Mor(V^*\tens W,\one)$ is zero if
$V\ncong W$ and 1 if $V\cong W$. This implies $T_{V^*\tens V}=0$ in
the former case. In the latter we have $\dim(\Mor(V^*\tens V,\one))=1$
and in the same way $\dim(\Mor(\one, V^*\tens V))=1$. This determines
a one-dimensional space of morphisms $V^*\tens V\to V^*\tens V$ to
which $T_{V^*\tens V}$ must belong. On the other hand,
$\coevi_V\circ\ev_V$ is an element of this space as well. Furthermore
it is non-zero as it can be converted to $\id_{V\tens V^*}$ by
suitable composition with $\coev_V$ and $\evi_{V^*}$. Thus, there
exists a complex number $\lambda$ such that $T_{V^*\tens V}= \lambda
\coevi_V\circ\ev_V$. Composing on both sides with $\ev_V$ yields
$\ev_V=\lambda \cdim_- V \ev_V$. As $\ev_V$ is non-zero this implies
$\cdim_- V\neq 0$ and furthermore $\lambda = (\cdim_- V)^{-1}$.

The statements for $V$ and $V^*$ interchanged follow correspondingly.
\end{proof}

\begin{figure}
\begin{center}
\input{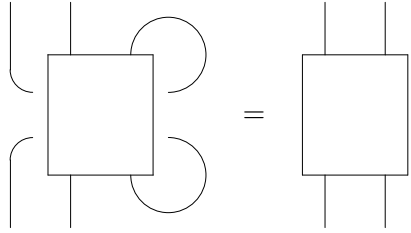}
\end{center}
\caption{Permutation identity for the $T$-morphism on a tensor
product.}
\label{fig:cyclT}
\end{figure}

\begin{prop}
Let $\catmod$ be a semisimple spherical category.
The \emph{permutation identity} for the $T$-morphism on a tensor product
depicted in Figure~\ref{fig:cyclT} holds. (The object labels and
arrows on the lines are arbitrary.)
\end{prop}
\begin{proof}
Choose decompositions for the two objects. Using naturality of $T$ the
identity can be reduced to an identity for the simple objects in the
decomposition. Thus, we have a tensor product of two simple objects
and we can apply Proposition~\ref{prop:dualid}. For the non-zero
contributions we
use the identity for $T_{V^*\tens V}$ on one side of
Figure~\ref{fig:cyclT} and the one for
$T_{V\tens V^*}$ on the other. As $\cdim_+=\cdim_-$ in a spherical
category we obtain equality.
\end{proof}

\section{Representation Theory}
\label{sec:reptheo}

In this section we review how the different types of categories arise
as categories of representations of groups, supergroups and various
types of Hopf algebras. We consider the issue of semisimplicity in
this context. Furthermore, we develop the necessary graphical notation
to represent functions on a group, supergroup or quantum group.

\subsection{Groups}
\label{sec:groups}

A relevant reference for Lie groups (representation theory, Haar
measure, Peter-Weyl decomposition) is e.g.\
\cite{CaSeMa:liegroups}.
Throughout this section, let $G$ be a group.

\subsubsection{Representation Categories}

In the following we
consider the category
of representations of $G$, which provides the most important example
of a symmetric category. We denote the action of a group element $g$ on a
vector $v$ by $g\act v$.

\begin{prop}
\label{ex:greptheo}
The category $\rep{G}$ of finite dimensional (left) representations of
a group
$G$ together with their intertwiners is a symmetric
category in the following way:
\begin{itemize}
\item
The monoidal structure is given by the tensor product of
representations. That is, for two representation $V,W$ we have a
representation $V\tens W$ via
\[
 g\act(v\tens w) \defeq g\act v \tens g\act w.
\]
The unit object $\one$ is the trivial representation.
\item
The rigid structure is given by the dual $V^*$ of a
representation $V$. This is the dual vector space with the action
\[
 \langle g\act f, v\rangle \defeq \langle f, g^{-1}\act v\rangle
\]
for all $g\in G, v\in V, f\in V^*$.
$\ev_V$ is simply the pairing between $V^*$ and $V$ while
$\coev_V:1\mapsto \sum_i v_i\tens f^i$ where $\{v_i\}$ is some basis of
$V$ and $\{f^i\}$ is the corresponding dual basis of $V^*$.
\item
The symmetric structure is given by the trivial braiding
\[
 \psi_{V,W}(v\tens w)= w\tens v .
\]
\end{itemize}
The simple objects in $\rep{G}$ are the irreducible
representations of $G$.
\end{prop}

We have now the diagrammatic formalism of
Section~\ref{sec:diag} at our disposal for group representations
and their intertwiners.

\subsubsection{Representative Functions}
\label{sec:gfunc}

We shall be particularly interested in the diagrammatic
representation of functions on the group considered as
representations. We discuss
this in the following.

Let $\falg_{alg}(G)$ denote the complex valued representative
functions on $G$. These
are the functions that arise as matrix elements of finite-dimensional
complex representations of $G$. That is, any representative function
is of the form
\begin{equation}
 g\mapsto \langle \phi, \rho_V(g) v\rangle ,
\label{eq:repfn}
\end{equation}
where $V$ is some finite-dimensional representation,
$\rho_V$ denotes the representation matrix and $v\in V,
\phi\in V^*$. 
We can thus identify the function with the vector $\phi\tens v$ in
$V^*\tens V$.
The sum of two representative functions is again a
representative function by the identity
\begin{equation}
 \langle \phi, \rho_V(g) v\rangle
 + \langle \phi', \rho_{V'}(g) v' \rangle
 = \langle \phi+\phi', \rho_{V\oplus V'}(g) (v+v') \rangle
\label{eq:addfunc}
\end{equation}
for the direct sum of representations. Similarly for the
product
\begin{equation}
 \langle \phi, \rho_V(g) v\rangle \cdot
 \langle \phi', \rho_{V'}(g) v' \rangle
 = \langle \phi\tens \phi', \rho_{V\tens V'}(g) (v\tens v') \rangle
\label{eq:multfunc}
\end{equation}
by the tensor product of representations.

\begin{figure}
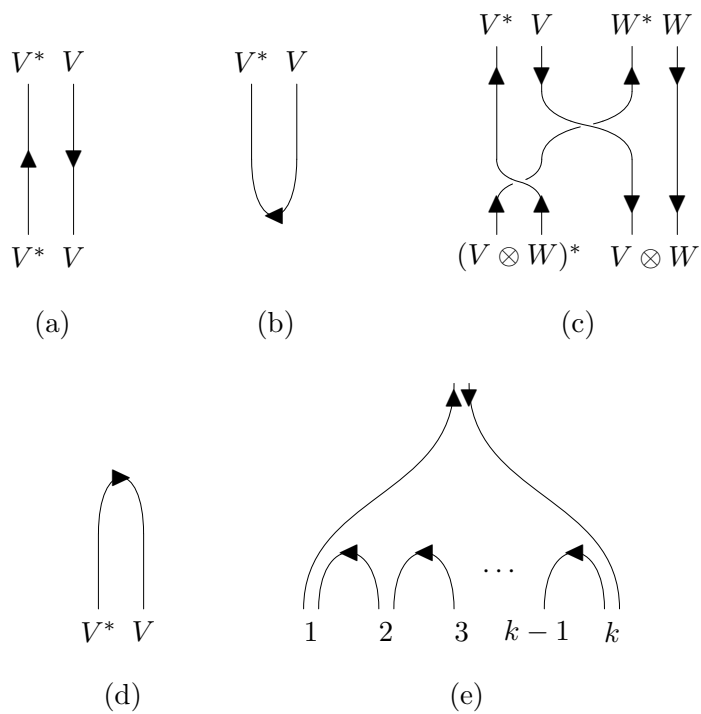

\begin{center}
\begin{tabular}{cp{1cm}cp{1cm}c}
\input{figures/fig_dline} && \input{figures/fig_counit} &&
\input{figures/fig_prodfunc}\\
\\
(a) && (b) && (c)
\end{tabular}
\\ \vspace{5mm}
\begin{tabular}{cp{1cm}c}
\input{figures/fig_char} && \input{figures/fig_copfunc} \\
\\
(d) && (e)
\end{tabular}
\caption{(a) Double line diagram for representative function.
(b) Evaluation at the group identity.
(c) Multiplication of representative functions.
(d) A character.
(e) Expansion of a representative function on a product of group
elements.}
\label{fig:func}
\end{center}
\end{figure}

Consider the action of $G$ on its algebra of functions by
conjugation as 
\begin{equation}
 (g\act f)(h)\defeq f(g^{-1} h g) .
\label{eq:conjact}
\end{equation}
As we will see in Section~\ref{sec:gaugesym} this action is
intimately related to gauge
transformations in lattice gauge theory.
For a representative function we have the identity
\begin{equation}
 (h\act (\phi \tens v))(g)=
 (h\act \phi \tens h\act v)(g)=
 (\phi \tens v)(h^{-1} g h) .
\end{equation}
That is, this action by conjugation is just the same thing as the
action on $V^*\tens V$ considered as a tensor product of
representations.
Consequently, we can denote a representative function
diagrammatically by a double line, one for $V$ and one for $V^*$,
see Figure~\ref{fig:func}.a.
In the following we consider arbitrary elements of $V^*\tens V$ as
representative functions so that besides the addition by
direct sum (\ref{eq:addfunc}) we also have the vector addition inside
$V^*\tens V$.
By definition,
the evaluation of a function at the group identity is just the
evaluation of the pairing, see Figure~\ref{fig:func}.b.
As the multiplication is given by the tensor product
(\ref{eq:multfunc}) it can be diagrammatically represented as in
Figure~\ref{fig:func}.c.

A type of function that is of particular importance in lattice gauge
theory is the character $\chi_V$ of a representation $V$. As an
element of $V^*\tens V$ it is
\begin{equation}
 \chi_V=\sum_n \phi_n\tens v_n ,
\label{eq:char}
\end{equation}
where $\{v_i\}$ denotes a basis of $V$ and $\{\phi_i\}$ a dual basis
of $V^*$.
Diagrammatically, this is an (arrow-reversed) coevaluation, see
Figure~\ref{fig:func}.d.
The invariance of a character under conjugation is reflected
by the fact that its diagram is closed to the top.
Note that the constant function with value 1 is the character for the
trivial representation.

Evaluating a representative function on a product of group elements
yields the expansion
\begin{equation}
  (\phi\tens v)(g_1\cdots g_k) 
  =\sum_{n_1,\dots,n_{k-1}} (\phi\tens v_{n_1})(g_1)
 (\phi_{n_1}\tens v_{n_2})(g_2)\cdots (\phi_{n_{k-1}}\tens v)(g_k) .
\label{eq:copfunc}
\end{equation}
Diagrammatically this expansion is the insertion of coevaluation
diagrams, see Figure~\ref{fig:func}.e.

\subsubsection{Integration and Semisimplicity}

If all finite-dimensional representations of $G$ are completely
reducible the category $\rep{G}$ is
semisimple and a normalized bi-invariant integral in the following
sense exists.

\begin{dfn}
\label{def:gint}
A normalized bi-invariant integral on $\falg_{alg}(G)$ is a map
$\int:\falg_{alg}(G)\to\C$ denoted $f\mapsto \int \xd g\, f(g)$
such that
\[
\int \xd g\,f(gh)=\int \xd g\,f(hg) = \int \xd g\, f(g)
\quad\forall h\in G
\quad\text{and}\quad \int \xd g=1 .
\]
\end{dfn}

Furthermore, this integral precisely defines the family of morphisms
$T$ of Proposition~\ref{prop:proj} in the following way.

\begin{prop}
For a representation $V$ the
intertwiner $T_V:V\to V$ of Proposition~\ref{prop:proj}
is given by the bi-invariant normalized integral as
\[
 T_V:v\mapsto \int\xd g\,\rho_V(g) v .
\]
\end{prop}

\begin{figure}
\begin{center}
\input{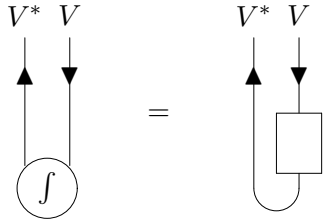}
\caption{Diagrammatic identity for the integral.}
\label{fig:integral}
\end{center}
\end{figure}

\begin{figure}
\begin{center}
\includegraphics{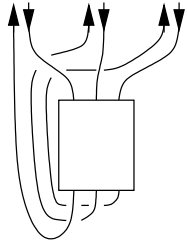}
\end{center}
\caption{The integral of a product of functions.}
\label{fig:prodint}
\end{figure}

Thus, the integral of a representative function is
\begin{equation}
 \int \xd g\, \langle \phi, \rho_V(g) v\rangle
 = \langle \phi, T_V(v) \rangle .
\label{eq:intf}
\end{equation}
Translating this formula into a diagram yields the identity shown in
Figure~\ref{fig:integral}. By combining this with
Figure~\ref{fig:func}.c we obtain Figure~\ref{fig:prodint} as the
diagrammatic representation of taking the integral of a product of
functions.

Two types of groups giving rise to semisimple representation
categories are of particular interest: compact Lie groups and
finite groups.

\begin{prop}[Peter-Weyl Decomposition]
\label{prop:gpw}
Let $G$ be a compact Lie group or a finite group. Then, $\rep{G}$ is
semisimple and the algebra of
representative functions on $G$ has a decomposition
\[
\falg_{alg}(G)\cong \bigoplus_V (V^*\tens V),
\]
where the direct sum runs over all irreducible representations $V$ of
$G$. The isomorphism is an isomorphism as representations of
$G\times G$ with the action $((g,g')\act f)(h)=f(g^{-1} h g')$ on the
left hand side and the canonical action on the right hand side.

The unique normalized bi-invariant integral $\int:\falg_{alg}(G)\to\C$
is given by the projection
\[
\bigoplus_V (V^*\tens V) \to \one^*\tens \one\cong\C,
\]
where $\one$ denotes the trivial representation.

In the Lie group case the representative functions are dense in the
$L^2$-functions of $G$, to which the integral (Haar measure) extends.

In the finite group case the integral can be expressed through a sum
\[
\int \xd g\, f(g)=\frac{1}{|G|}\sum_{g\in G} f(g),
\]
where $|G|$ denotes the order of $G$.
\end{prop}

\subsection{Hopf Algebras and Quantum Groups}

Hopf algebras can be considered as generalizations of groups in the
sense that they are ``noncommutative algebras
of functions''.
Hence the name \emph{quantum groups}.
The coproduct thereby encodes the ``group structure''.
We consider Hopf
algebras with various amounts of additional structure so that
their respective representation theory gives rise to all the
different types of categories we are interested in here.
See \cite{Maj:qgroups} for a general reference covering most of the
relevant cases. Spherical Hopf algebras are considered in
\cite{BaWe:sphcat}.

We use here the point of view that representations are
comodules. This is precisely in the spirit of ``noncommutative
function algebras'' and indeed, as we shall see below, a group is then
just (equivalent to) a certain Hopf algebra. This is also the right
point of view for supergroups, see e.g.\ \cite{Oe:qgeosusy}.
Examples of Hopf algebras giving rise
to nonsymmetric categories are then the $q$-deformations of simple
Lie groups. Dually, one can consider modules as representations. This
corresponds then to Hopf algebras generalizing universal
enveloping algebras.
However, this implies the loss of ``global
structure'' of the group.
But as this point of view is more frequently employed in the
literature, many of our definitions have a ``co'' in them.
In the case of finite-dimensional Hopf algebras both points of view
are completely equivalent. 

We use the notation $\cop, \cou, \antip$ for coproduct, counit and
antipode of a Hopf algebra. We use Sweedler's notation (with implicit
summation) $\cop a = a\i1 \tens a\i2$ for coproducts and a similar
notation $v\mapsto v\iu1 \tens v\i2$ for right coactions.

\subsubsection{Representation Categories}

\begin{dfn}
Let $H$ be a Hopf algebra and
$\omega:H\to\C$ a convolution-invertible map
such that
\[
 \omega(a b)=\omega(a)\omega(b),\quad \antip^2 a = \omega(a\i1) a\i2
 \omega^{-1}(a\i3)
\]
for all $a,b\in H$. We call $(H,\omega)$ a \emph{copivotal Hopf
algebra}.
\end{dfn}

\begin{dfn}
Let $(H,\omega)$ be a copivotal Hopf algebra. It is called a
\emph{cospherical Hopf algebra} if
for all right $H$-comodules $\beta:V\to
V\tens H$ and all comodule maps
$\theta:V\to V$ the equality
\[
\tr((\id_V\tens\omega)\circ\beta\circ\theta) =
\tr((\id_V\tens\omega^{-1})\circ\beta\circ\theta)
\]
holds.
\end{dfn}

\begin{dfn}
\label{def:cqtr}
A \emph{coquasitriangular structure} on a Hopf algebra $H$ is a
convolution-invertible map $\cR:H\tens H\to\C$ so that
\begin{gather*}
\cR(ab\tens c)=\cR(a\tens c\i1) \cR(b\tens c\i2),\quad
\cR(a\tens bc)=\cR(a\i1\tens c) \cR(a\i2\tens b),\\
b\i1 a\i1 \cR(a\i2\tens b\i2)=\cR(a\i1\tens b\i1) a\i2 b\i2
\end{gather*}
for all $a,b,c\in H$. A pair $(H,\cR)$ is called a
\emph{coquasitriangular Hopf algebra}.
\end{dfn}

\begin{dfn}
\label{def:ribform}
A \emph{ribbon form} on a 
coquasitriangular Hopf algebra $(H,\cR)$ is a map
$\nu:H\to\C$ such that
\begin{gather*}
 \nu(a b)=\cR^{-1}(a\i1\tens b\i1) \cR^{-1}(b\i2\tens a\i2)
 \nu(a\i3) \nu(b\i3),\\
 \nu(1)=1,\quad \nu(\antip a)=\nu(a),\quad
 \nu(a\i1) a\i2= a\i1 \nu(a\i2)
\end{gather*}
for all $a,b\in H$. A triple $(H,\cR,\nu)$ is called a \emph{coribbon
Hopf algebra}.
\end{dfn}

\begin{lem}
A coribbon Hopf algebra is a cospherical Hopf algebra by
setting $\omega(v)\defeq\cR^{-1}(\antip v\i1\tens v\i2)\,\nu(v\i3)$.
\end{lem}

\begin{dfn}
\label{def:ctr}
A \emph{cotriangular structure} on a Hopf algebra $H$ is
coquasitriangular structure satisfying the extra property
\begin{gather*}
\cR(a\tens b)=\cR^{-1}(b\tens a)
\end{gather*}
for all $a,b\in H$. A pair $(H,\cR)$ is called a
\emph{cotriangular Hopf algebra}.
\end{dfn}

\begin{lem}
A cotriangular Hopf algebra is a coribbon Hopf algebra by
choosing $\nu\defeq\cou$.
\end{lem}

\begin{prop}
\label{prop:hopfrep}
The category $\comod{H}$ of finite dimensional (right) comodules of a
Hopf algebra $H$ is a rigid monoidal category in
the following way:
\begin{itemize}
\item
The monoidal structure is given by the tensor product of comodules.
Thus, for two comodules $V,W$ we have a comodule structure on $V\tens
W$ via
\[
 v\tens w\mapsto v\iu1 \tens w\iu1\tens v\i2 w\i2 .
\]
The unit object $\one$ is the 1 dimensional trivial comodule $v\mapsto
v\tens 1$.
\item
The rigid structure is given by the definition of the dual $V^*$ of a
comodule $V$. This is the dual vector space with the coaction
\[
 f\mapsto f\iu1\tens f\i2\quad \text{such that}\quad
 \langle f\iu1, v\rangle f\i2 = \langle f, v\iu1\rangle \antip v\i2
\]
for all $v\in V, f\in V^*$.
$\ev_V$ is simply the pairing between $V^*$ and $V$ while
$\coev_V:1\mapsto \sum_i v_i\tens f^i$ where $\{v_i\}$ is some basis of
$V$ and $\{f^i\}$ is the corresponding dual basis of $V^*$.
\item
If $H$ is copivotal/cospherical, then $\comod{H}$ is a pivotal/spherical
category by
defining $\tau_V:V\to V^{**}$ for an object $V$ as
\[
 \tau_V:v\mapsto v\iu1\,\omega(v\i2),
\]
where $V$ and $V^{**}$ are identified canonically as vector spaces and
the coaction is the one on $V$.
\item
If $H$ is coribbon, then
$\comod{H}$ is a ribbon category with braiding
\[
 \psi_{V,W}(v\tens w)= w\iu1\tens v\iu1 \cR(v\i2\tens w\i2) .
\]
and twist
\[
 \nu_V:v\mapsto v\iu1 \,\nu(v\i2) .
\]
\item
If $H$ is cotriangular, then the category is symmetric with the
braiding obtained as
\[
 \psi_{V,W}(v\tens w)= w\iu1\tens v\iu1 \cR(v\i2\tens w\i2) .
\]
\end{itemize}
\end{prop}

To see how the group case is manifestly a special case of the
cotriangular Hopf algebra case note the following fact.

\begin{prop}
\label{prop:grphopf}
Let $G$ be a group. Then $\falg_{alg}(G)$ is naturally a commutative
Hopf algebra. The coproduct is given by the map
\[
 V^*\tens V\to (V^*\tens V)\tens (V^*\tens V):
 \phi\tens v\mapsto \sum_n (\phi\tens v_n)\tens (\phi_n\tens v)
\]
and the antipode is given by
$V^*\tens V\to V\tens V^*:\phi\tens v\mapsto v\tens \phi$, using that
canonically $V^{**}\cong V$ as representations.

Furthermore, a finite-dimensional representation of $G$ is canonically
the same thing as a finite-dimensional comodule of $\falg_{alg}(G)$
and vice versa by the coaction
\[
 V\to V\tens (V^*\tens V): v\mapsto \sum_n v_n\tens (\phi_n\tens v) .
\]

If $\falg_{alg}(G)$ is equipped with the trivial cotriangular
structure $\cR=\cou\tens\cou$, then $(H,\cR)$ can be said to
correspond to $G$ in the sense that
$\comod{H}$ is identical to $\rep{G}$.
\end{prop}

The natural definition of a supergroup in view of the
above proposition is that of
$\Z_2$-graded commutative Hopf algebra. The
elements of the Hopf algebra play the role now of ``functions on
the supergroup''.
A $\Z_2$-graded Hopf algebra satisfies the same axioms as a Hopf
algebra, except for the compatibility of product and coproduct which
is modified to
\begin{equation}
 \cop(a b)= (-1)^{|a\i2| |b\i1|} a\i1 b\i1 \tens a\i2 b\i2 .
\end{equation}
$\Z_2$-graded commutativity means then $a b = (-1)^{|a| |b|}=b a$.
To a $\Z_2$-graded commutative Hopf algebra corresponds precisely a
cotriangular Hopf algebra which is obtained from the former one by
``bosonization'' and has exactly the same
representation theory \cite{Maj:qgroups}.

However, the cotriangular Hopf algebra point of view is more general
and superior from
a physical point of view as it allows the algebraic implementation of
spin-statistics relations. In particular, this is relevant for the
formulation of supersymmetric theories.
The convenient definition of supergroup in our context is thus that
of a cotriangular Hopf algebra $H$ equipped with a surjection to the
Hopf algebra of functions on $\Z_2$ with nontrivial braiding.
See \cite{Oe:qgeosusy} for details.
The cotriangular structure encodes the $\Z_2$-grading on the
representations (comodules) as
\begin{equation}
\psi_{V,W}(v\tens w)= (-1)^{|v| |w|} w\tens v .
\end{equation}

\subsubsection{``Representative Functions''}

We now consider how elements of a Hopf algebra can be dealt
with in a similar manner as representative functions on a group.

Let $H$ be a Hopf algebra.
For a finite-dimensional vector space $V$ the space $V^*\tens V$ has
canonically the structure of a coalgebra by the coaction
\begin{equation}
\cop \phi\tens v=\sum_n (\phi\tens v_n)\tens (\phi_n\tens v)
\quad\text{and counit}\quad
\cou(\phi\tens v)=\langle \phi, v\rangle .
\label{eq:comcop}
\end{equation}
Given a finite-dimensional comodule $V$ of $H$ we obtain a map of
coalgebras $V^*\tens V\to H$ via
\begin{equation}
 \phi\tens v\mapsto \langle \phi, v\iu1\rangle v\i2 .
\label{eq:qmap}
\end{equation}
In fact, any element of $H$ is in the image of such a map. To see this,
consider $H$ as a right comodule under itself by the coproduct.
Any given $h\in H$ is contained in some finite-dimensional subcomodule
$V\subseteq H$. Choose $\phi\in V^*$ such that $\phi(v)=\cou(v)$ for
any $v\in V$. Then, $\phi\tens h\mapsto h$ under the above map.

Therefore, similarly to the group case, we can represent elements of
$H$ by double line diagrams.
The coaction of $H$ on itself implicit
in this notation is now the right adjoint coaction
$h\mapsto h\i2\tens (\antip h\i1) h\i3$.
The diagram for the counit is just the
evaluation.
The diagram for the coproduct
is obtained by inserting coevaluations as follows from
(\ref{eq:comcop}). To make the analogy with the group case complete,
we can define a character by the (arrow-reversed) coevaluation
diagram. We obtain then exactly the diagrams (a), (b), (d), (e) of
Figure~\ref{fig:func}. Note that the counit expresses evaluation at
the identity while the coproduct expresses
evaluation on a product of group elements in the group case.

\subsubsection{Integration and Semisimplicity}

Proceeding in an analogous way as for groups, 
we introduce in the semisimple case the integral which defines the
$T$-morphism and its diagrammatic representation.

\begin{dfn}
\label{def:qint}
Let $H$ be a Hopf algebra. A bi-invariant normalized integral on $H$
is a map $\int: H\to\C$ such that
\[
 h\i1 \int h\i2 = \left(\int h\i1\right) h\i2 
= 1 \int h \quad\forall h\in H \quad\text{and}\quad \int 1=1 .
\]
\end{dfn}
Note in particular that Definition~\ref{def:gint} is the special case
of Definition~\ref{def:qint} for $H=\falg_{alg}(G)$.

\begin{prop}
For an object $V$ the intertwiner $T_V:V\to V$ of
Proposition~\ref{prop:proj} is given by the bi-invariant normalized
integral as
\[
 T_V: v\mapsto v\iu1 \int v\i2 .
\]
\end{prop}

Note that in the double line diagrammatics considered above we obtain
the identity of Figure~\ref{fig:integral} as from (\ref{eq:qmap}) we
have
\begin{equation}
 \int (\phi\tens v) \mapsto \langle \phi, v\iu1\rangle \int v\i2
 = \langle \phi, T_V(v)\rangle .
\end{equation}

The generalization of a compact Lie group or finite group is a
cosemisimple Hopf algebra.
That is, a Hopf algebra which is as a coalgebra a direct sum of
simple coalgebras.
The structure of cosemisimple Hopf algebras is captured by the
following proposition.

\begin{prop}[Peter-Weyl Decomposition]
\label{prop:pw}
Let $H$ be a cosemisimple Hopf algebra. Then $\comod{H}$ is semisimple
and the following isomorphism of coalgebras holds
\[
H\cong \bigoplus_V (V^*\tens V),
\]
where the direct sum runs over all simple (right) comodules $V$ of $H$.
$V^*\tens V$ is a simple coalgebra as above with the isomorphism as
given there.

The unique normalized left and right invariant integral $\int:H\to\C$
is given by the projection
\[
\bigoplus_V (V^*\tens V) \to \one^*\tens \one\cong\C,
\]
where $\one$ denotes the trivial comodule.
\end{prop}

\section{The Partition Function}

We start by fixing some terminology.
In the following, \emph{complex} means finite CW-complex, see e.g.\
\cite{Mas:algtop}.
A \emph{lattice} means a finite combinatorial 2-complex. For a lattice
we use the terms
\emph{vertex, edge, face} to denote 0-, 1-, and 2-cells respectively.
We use the term \emph{cellular manifold} to denote a compact
manifold together with a cellular decomposition as a
finite CW-complex. The \emph{lattice associated with a cellular
manifold} means the 2-skeleton of the dual complex.

A standard reference for lattice gauge theory is \cite{Cre:lgt}.

\subsection{Ordinary and Symmetric LGT}
\label{sec:lgt}

It is well known that lattice gauge theory admits a spin foam
formulation \cite{Rei:worldsheet}. In
\cite{OePf:dualgauge} the corresponding transformation was explicitly
performed on a hypercubic lattice and it was shown that the new
formulation is strong-weak dual to the original one.
We perform here a generalization of this
transformation employing the categorial and diagrammatic
language introduced in the previous sections. 
This allows us to generalize LGT to arbitrary semisimple
symmetric categories (e.g.\ including supersymmetric LGT).

Let $L$ be a lattice and $G$ a compact Lie group or finite group. We
might think of $L$ as arising as the lattice associated with a cellular
manifold $M$. We equip the faces of $L$ with arbitrary but fixed
orientations.
Recall that in lattice gauge theory a group element $g$ is attached to
every edge (with a given orientation). The action $S$ is the sum over
all faces $f$ of a function $\sigma$ evaluated on the product
of group elements attached to
the edges $e$ bounding the face $f$ (cyclicly ordered by the
orientation of the face). We write this as
\begin{equation}
 S=\sum_f \sigma(\prod_{e\in \partial f} g_e) .
\label{eq:lgtact}
\end{equation}
The function $\sigma$ is required to be invariant under conjugation
(\ref{eq:conjact}) and to satisfy $\sigma(g^{-1})=\sigma(g)$.
The conjugation invariance ensures that it does not matter at which
vertex we start taking the product over group elements, only the
cyclic order is relevant.
The partition function reads
\begin{equation}
 \pf=\int (\prod_e \xd g_e)\, e^{-S}
  =\int (\prod_e \xd g_e) \prod_f
   e^{-\sigma(\prod_{e\in \partial f} g_e)} .
\label{eq:lgtpf}
\end{equation}
As $e^{-\sigma}$ is itself invariant under conjugation it can be
expanded as
\begin{equation}
 e^{-\sigma}=\sum_V \alpha_V \chi_V ,
\end{equation}
where the sum runs over the irreducible representations $V$ of $G$ and
$\chi_V$ is the character of the representation $V$. Since
$e^{-\sigma(g^{-1})}= e^{-\sigma(g)}$ we have $\alpha_{V^*}=\alpha_V$
because of $\chi_V(g^{-1})=\chi_{V^*}(g)$.
Thus
\begin{align}
 \pf & = \int (\prod_e \xd g_e) \prod_f \sum_V \alpha_V
 \chi_V(\prod_{e\in f} g_e)
  = \sum_{V_f} (\prod_f \alpha_{V_f})\, \pf_{V_f} 
\label{eq:partexp}\\
\text{with}\quad \pf_{V_f} & \defeq
 \int (\prod_e \xd g_e) \prod_f \chi_{V_f}(\prod_{e\in f} g_e) ,
\label{eq:ffunc}
\end{align}
where $V_f$ denotes an assignment of an irreducible representation
$V$ to every face $f$ and we sum over all such assignments.

The next step is to expand the characters into functions (i.e.\ matrix
elements)
taking as values the individual group elements attached to
the edges. Then, for each edge all the functions taking the attached
group element as their value are multiplied and integrated.
Instead of proceeding formally we perform these manipulations
diagrammatically. This leads to a diagrammatic representation of
the partition function. The obtained diagram is naturally embedded
into the lattice and into the manifold (if the lattice arises from
one).

To represent functions on the group diagrammatically we utilize the
diagrammatic language introduced in the previous sections. Notably,
we consider an action of the gauge group by conjugation
(\ref{eq:conjact}). As we
shall discuss in Section~\ref{sec:gaugesym} this precisely encodes
gauge transformations.

\begin{figure}
\begin{center}
\input{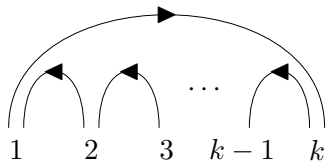}
\end{center}
\caption{The diagram of a character evaluated on a product of group
elements.}
\label{fig:copchar}
\end{figure}

\begin{figure}
\begin{center}
\includegraphics{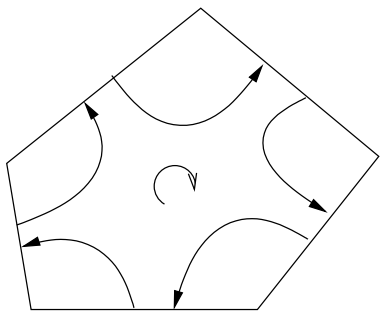}
\end{center}
\caption{The diagram of a character embedded into
a face.}
\label{fig:facechar}
\end{figure}

\begin{figure}
\begin{center}
\includegraphics{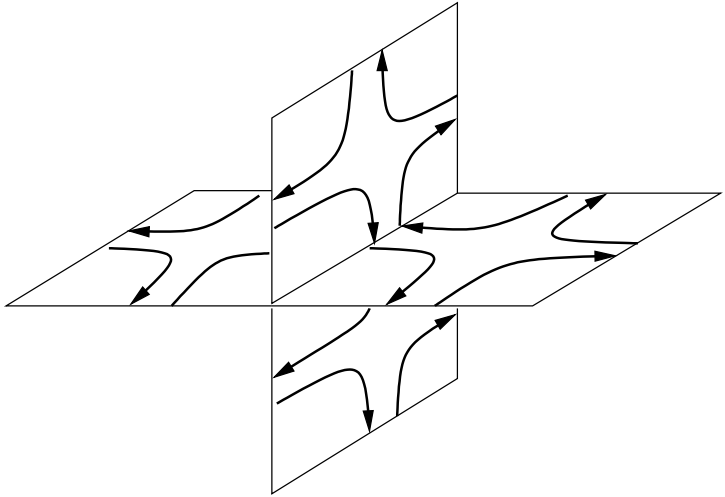}
\end{center}
\caption{
The character diagrams embedded into a 3-dimensional cubic
lattice. In this example four double lines meet at each edge.}
\label{fig:edgechars}
\end{figure}

As introduced in Section~\ref{sec:gfunc} a character $\chi_V$ is
represented by a coevaluation diagram (Figure~\ref{fig:func}.d) due to
its decomposition (\ref{eq:char}) as basis and dual basis over the
representation $V$. Thus, from (\ref{eq:ffunc}) we obtain one such
diagram for each face. As each character is evaluated on the product of
group elements attached to the edges bounding the face, we expand it
into functions of the individual group elements. This means attaching
the diagram of Figure~\ref{fig:func}.e to each character diagram. We
obtain a diagram as shown in Figure~\ref{fig:copchar} for every
face whereby the double lines correspond to the edges bounding the
face. Thus, we can embed the diagram for each face into the face as
shown in Figure~\ref{fig:facechar}.
The direction of the arrows is chosen in correspondence to the
orientation of the face.
Proceeding in this way for each face, all double lines
denoting functions that take the group value for a given edge meet in
this edge, see Figure~\ref{fig:edgechars}.

\begin{figure}
\begin{center}
\includegraphics{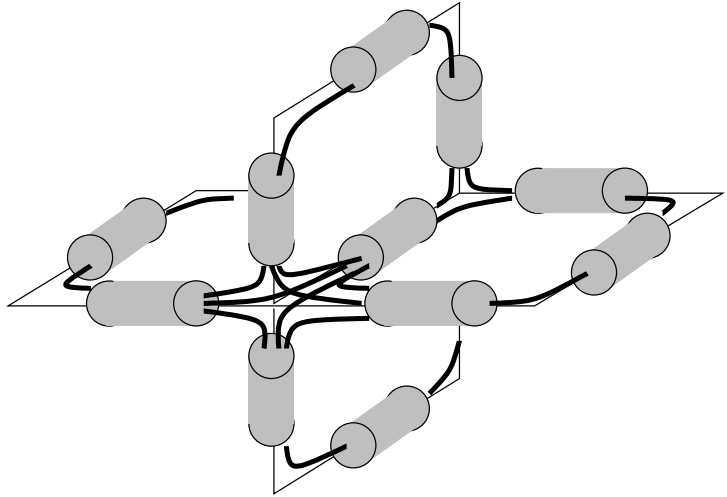}
\end{center}
\caption{
The circuit diagram obtained after inserting the
integrals. This example 
shows wires (thick lines) and cables (gray cylinders) in a piece
of 3-dimensional cubic lattice. The
arrows on the wires are omitted.}
\label{fig:edgecab}
\end{figure}

To integrate the product of functions taking their value at a given
edge as prescribed by (\ref{eq:ffunc}) we insert the appropriate
diagram (Figure~\ref{fig:prodint}) at each edge, connecting the double
lines.
For ease of notation, we now draw the lines in each face such that they
run close to the boundary. We draw
the $T$-projections (the unlabeled coupon in
Figure~\ref{fig:prodint}) arising in the integration as (hyper)-cylinders
with axis given by the edges. Proceeding thus for every edge we
arrive at a diagram embedded in the lattice as shown in
Figure~\ref{fig:edgecab}.

As this diagram is closed it represents an intertwiner $\one\to\one$
and thus a complex number which is exactly $\pf_{V_f}$
in (\ref{eq:ffunc}). Thus,
the partition function is the sum over this diagram with all
possible assignments of irreducible representations to the faces with
weights given by the $\alpha_V$. Note that this is independent of the
choice of orientations for the faces. Indeed, changing the orientation
of a face leads to the same partition function, except that
$\alpha_V$ and $\alpha_{V^*}$ for this face are interchanged. However,
they are equal by assumption. 

Since the diagram resembles an arrangement
of wires and cables we denote the lines representing the characters as
\emph{wires} and the (hyper)-cylinders representing the $T$-projectors
as \emph{cables}. When referring to an individual wire we
usually mean all the lines lying in a given face as they come from the
same character. They carry the same representation label and arrow
orientation and
we can imagine them being connected inside the cables.

Let us remark that the arrows on the wires in a given cable do not
necessarily all
point in the same direction (as in Figure~\ref{fig:prodint}). This
indeed must be so as those functions originating from faces with opposite
orientation with respect to the one in which the edge carries the group
value $g$ are evaluated at $g^{-1}$ instead.
The relative directions of the arrows encode this information.

Starting from the lattice the diagram is obtained in a very simple
way: Put one wire into each face running close to the boundary. Give
each wire an arrow according to the orientation of
the face and the representation label of the face. Then, for each
edge, put a cable in the middle of the edge, around the wires that run
along the edge. We refer to the diagram as the \emph{circuit diagram}
associated to the lattice. Note that it is a spin network in the sense
of Section~\ref{sec:diag}.

Strictly speaking, we have in Section~\ref{sec:diag} only defined
how to
evaluate a diagram that can be written on a ``piece of paper'', i.e.,
in the plane. However, as discussed there, the value of the diagram only
depends on the
combinatorial data, i.e., which piece of wire is connected to which
side of which cable and with which arrow. This information is
completely determined by the lattice. Any way of writing the diagram
on a piece of paper (we refer to this as \emph{projection})
will give the same result.

Changing our point of view, we can now consider the obtained
representation of the partition function as a definition and thereby
extend it to arbitrary semisimple symmetric categories.

\begin{dfn}
\label{def:lgt}
Let $\catmod$ be a semisimple symmetric
category. Let $\{\alpha_V\}$ be an assignment of a complex number to
each isomorphism class of simple object $V$ such that
$\alpha_{V^*}=\alpha_V$. These are called \emph{weights}.
Let $L$ be a lattice, possibly associated with a cellular manifold $M$.
This defines a \emph{lattice gauge theory} as follows.

For any choice of orientation and labeling $V_f$ with an equivalence
class of simple objects for each face we define
$\pf_{V_f}$ to be the value of the circuit diagram constructed above.
We call
\[
 \pf\defeq\sum_{V_f} (\prod_f \alpha_{V_f}) \pf_{V_f}
\]
the \emph{partition function}, where the sum runs over all possible
labelings. This does not depend on the chosen orientations of the
faces.
\end{dfn}

By the above derivation, this definition agrees with ordinary lattice
gauge theory in the case where $\catmod=\rep{G}$.
Note that as $\pf_{V_f}$ is finite by construction, $\pf$ is manifestly
finite if $\catmod$ has only finitely many isomorphism classes of
simple objects. However, it might be infinite in general.
In ordinary lattice gauge theory it is finite by construction in spite
of the sum over labelings being infinite for a Lie group.

\subsection{Nonsymmetric LGT}
\label{sec:qlgt}

In the present section we extend our definition of LGT to
nonsymmetric categories.
This turns out to require topological
information beyond the lattice. That is, we need to start with a
cellular manifold and the admissible type of nonsymmetric category
depends on the dimension of the manifold. Also the manifold is now
required to be orientable. The type of category
admissible is the same as for state-sum invariants of Turaev-Viro
type, namely spherical in 3 dimensions \cite{BaWe:invplm}
and ribbon in 4 dimensions \cite{CrKaYe:inv4}.
This is not surprising as these invariants arise indeed as special
cases of our construction. This will be discussed in
Section~\ref{sec:topinv}.
In 2 dimensions we can use pivotal categories.

That the circuit diagram constructed in the previous section cannot
alone be used to define a partition function is clear. For pivotal,
spherical and ribbon categories the value of the diagram depends on
the way it
is projected onto the plane and not just on its combinatorial
data. This extra data is extracted from topological information
about the cellular manifold.

Let $M$ be an oriented cellular 2-manifold and consider its circuit
diagram. The only obstruction to its direct evaluation is the
non-planar topology of $M$. Instead, we cut out all the 2-cells (with
the wire pieces) and project them separately onto the plane, using the
orientation of $M$. This cuts all the cables in half. Now, we
reconnect the cables with $T$-coupons, thereby possibly introducing
crossings. The ``layout'' for these $T$-coupons is irrelevant however,
as they can arbitrarily cross (Figure~\ref{fig:Tcross}) and ``twist''
(Figure~\ref{fig:r1T}). Thus, we obtain a
well-defined morphism $\one\to \one$ (which is a complex number) in the
pivotal category.

The situation in higher dimensions is more involved and we start by
outlining the
$n$-dimensional setting before specializing to $n=3$ and $n=4$.

Let $M$ be an oriented cellular manifold of dimension $n$ and $L$
the associated lattice, embedded into $M$. The key idea is to embed
the circuit diagram into the $n-1$ dimensional subcomplex of $M$.
Again we choose an arbitrary orientation for each face (or
equivalently $(n-2)$-cell).

\begin{figure}
\begin{center}
\includegraphics{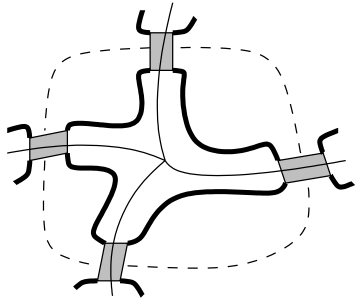}
\end{center}
\caption{The circuit diagram embedded into the $(n-1)$-complex. The
thin lines are the $(n-1)$-cells. The gray boxes are the cables. The
thick
lines are the wires, for illustration slightly displaced from the
boundaries into the $n$-cells. The face dual to the $(n-2)$-cell at the
meeting point in the center is indicated by a dashed line.}
\label{fig:cellcirc}
\end{figure}

Instead of embedding the wires arbitrarily into the faces of the
lattice we put them on the intersections of these faces with the
$(n-1)$-cells. Thus, every $(n-1)$-cell carries two pieces of wire for
each face that it intersects. The two correspond to the two (not
necessarily distinct) $n$-cells that are bounded by the
$(n-1)$-cell and they lie on top of each other. (It might be helpful
to imagine the wires to be slightly
displaced into the respective $n$-cells.)
Each piece of wire carries the object label and direction inherited from
the face. The wire pieces end precisely at the intersections of the
$(n-1)$-cells with the edges. Now we put a ``small'' $(n-2)$-sphere
around each intersection of an $(n-1)$-cell with an edge, into the
$(n-1)$-cell. We let the wire pieces end on those $(n-2)$-spheres
instead of the edges. The $(n-1)$-balls bounded by these
$(n-2)$-spheres are to be thought of as the (infinitely shortened)
cables. See Figure~\ref{fig:cellcirc} for an illustration.

\begin{figure}
\begin{center}
\includegraphics{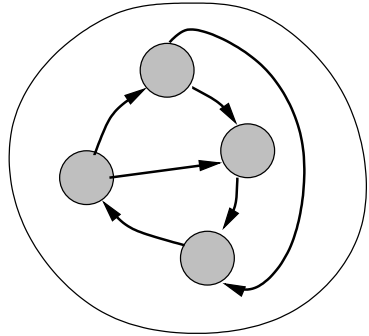}
\end{center}
\caption{The projection of the boundary of an $n$-cell with wires and
cables. The wire pieces end on the boundaries of the cables
($(n-1)$-balls) which are shaded.}
\label{fig:cellproj}
\end{figure}

\begin{figure}
\begin{center}
\includegraphics{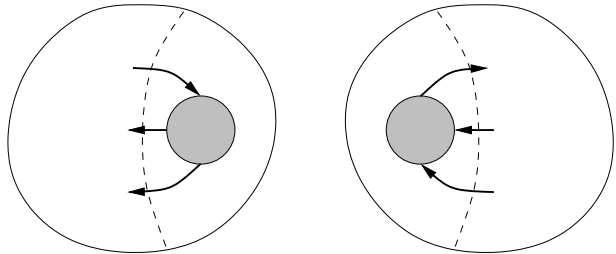}
\end{center}
\caption{Two projected $(n-1)$-spheres containing the same
$(n-1)$-cell. The projections of this $(n-1)$-cell (indicated by the
dashed line) are mirror images.}
\label{fig:2cellproj}
\end{figure}

\begin{figure}
\begin{center}
\includegraphics{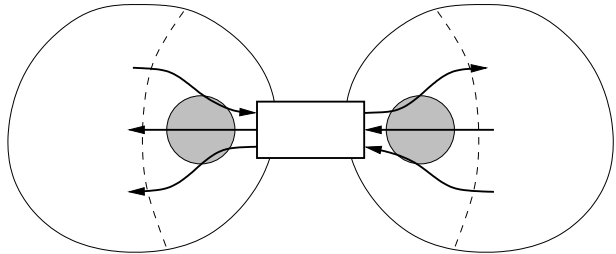}
\end{center}
\caption{Reconnecting the cable ends (gray disks) by a $T$-coupon.}
\label{fig:pconnect}
\end{figure}

By definition of a CW-complex, we can think of the boundary of an
$n$-cell always as an
$(n-1)$-sphere with some of its constituent $(n-1)$-cells possibly
identified. It is natural to consider this $(n-1)$-sphere (before
identification) as carrying the wire pieces belonging to the
$n$-cell.
The idea for evaluating the circuit diagram is now to ``cut out'' the
$(n-1)$-sphere for each $n$-cell with its wire pieces and to project
it onto the plane to define (almost) a morphism (upon labeling), see
Figure~\ref{fig:cellproj}. Note that each $(n-1)$-cell
occurs twice in the projections, once for the $n$-cell that it
bounds on each side. (These $n$-cells might be identical leading to
the $(n-1)$-cell appearing twice in one projected $(n-1)$-sphere.)
Using the orientation in performing the projections the two occurrences
of each $(n-1)$-cell agree in the sense of being mirror images, see
Figure~\ref{fig:2cellproj}.
Then one reconnects all the individual diagrams at the matching
(mirror image) cable ends with $T$-coupons, see
Figure~\ref{fig:pconnect}. We call
this the \emph{projected circuit
diagram} which (upon labeling) gives rise to a morphism $\one\to\one$
in the relevant category and thus a complex number.

Our description already suggests that the admissible type of category
depends on the isotopy properties of the diagrammatics. More
precisely, we should have isotopy invariance on $S^{n-1}$, as this is
where the (unconnected) diagrams live.
Indeed, this approach can be realized in dimension 3 for spherical
categories and in dimension 4 for ribbon categories, compare
Table~\ref{tab:cat}. We describe this in the following.

\subsubsection{Dimension 3 - Spherical Categories}
\label{sec:sphlgt}

Before projecting, we mark each of the circles that represent the
cables at some arbitrary point (not coinciding with the end of a wire
piece). Then,
to perform the projections, each 2-sphere (bounding a 2-cell) is
punctured at some arbitrary point (which does not lie in any of the
circles or on any of the wire pieces). Now, each 2-sphere is flattened
out and projected onto the plane
such that its outside is facing up. In doing so we respect the
orientation of $M$. This ensures that the orientations of the
projected 2-spheres all match up in the sense that the two projections
of each 2-cell are mirror images. We wish to think of the images of
the 2-spheres as diagrams defining (upon labeling with objects)
morphisms in the category. For each projected cable end (represented
by a circle) we arrange the wires ending there in a line by cutting
the circle at the marked point. Now if we could pull these lines to
the top or bottom line of the diagram this would (upon labeling)
specify a morphism. As this would possibly imply introducing crossings
this is not in general possible. Nevertheless, we can connect the
corresponding wires lined up at the cable ends by $T$-coupons. This is
because $T$-coupons are allowed to cross arbitrarily. The emerging
diagram is the desired projected circuit diagram. It defines (upon
labeling) a morphism $\one\to\one$ in the spherical category and thus
a complex number.

We proceed to show the well definedness (invariance under the choices
made) of the obtained morphism (for
any labeling). For given projections, the morphism is independent on
the way the $T$-coupons are inserted to connect matching cable
ends. This is because the $T$-coupons can cross arbitrarily
(Figure~\ref{fig:Tcross}) and ``windings'' in the connections are
irrelevant (Figure~\ref{fig:r1T}).
The invariance of the morphism under the way the projections of the
2-spheres are performed (while respecting the orientation) follows
precisely from the $S^2$ isotopy invariance of the
diagrammatics. Notably, the invariance under the choice of point at
which each boundary 2-sphere is pierced is precisely the identity of
Figure~\ref{fig:sphprop}. Note that a diagram that is bounded only by
cables behaves in this sense like a closed diagram. It remains to show
the invariance under the choice of marked point for each cable. Moving
this point across the end of a wire gives precisely rise to diagrams
that are related as the sides of Figure~\ref{fig:cyclT}. As these are
identical invariance follows.

\subsubsection{Dimension 4 - Ribbon Categories}
\label{sec:riblgt}

We start by turning the wires into ribbons, i.e.,
equip them with a framing. As we confine the framing to the boundary
3-spheres in which the wires lie this gives indeed rise to ribbons
(again inside the 3-spheres).
It turns out that one is free to choose
the framing as long as it is coherent in the following way. Recall
that the wire pieces all occur in pairs (each belonging to one of the
two bounded 4-cells) which lie on top of each other. 
A framing
is said to be coherent, if for each piece of wire the corresponding
piece of wire has exactly the same framing, but face-side and backside
exchanged.
Next, we arrange the ribbon wire endings on each of the 2-sphere
cables in a line, with the framings pointing along the
line, all faces (for each bounded 4-cell) on the same side.
We do this while keeping corresponding wire pieces identified (except
for their face-side and backside being opposite).

Now, we puncture each of the bounding 3-spheres at a point to identify
them with $\R^3$. Hereby we respect the orientation and choose the
``outward'' direction for all the 3-spheres to be the same in the
ambient $\R^4$ (considering the $\R^3$ as a subspace of this).
(This is the analog to projecting 2-spheres ``face up'' in the 3
dimensional case.)
Then, for each 3-sphere we project the obtained ribbon tangle
in $\R^3$ onto the plane. We do this in such a way that the aligned
ribbons that end on the 2-spheres (cables) are projected
``face up''.
To let the projections define morphisms (upon labeling) we would need
to pull the ribbon ends to the top or bottom line of each diagram.
However, we do not need to do that but can proceed to connect the
corresponding ribbon ends with $T$-coupons. The
resulting diagram is the desired projected circuit diagram.
It defines (upon labeling) a morphism $\one\to\one$ in
the category and thus a complex number.

\begin{figure}
\begin{center}
\input{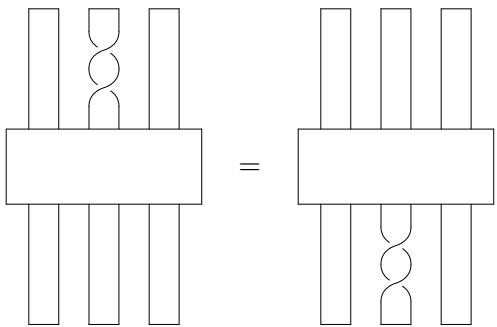}
\end{center}
\caption{Pulling a twist through a $T$-coupon.}
\label{fig:twistcom}
\end{figure}

We proceed to show the well definedness (invariance under the choices
made) of the obtained morphism (for
any labeling). For given projections, the morphism is independent on
the way the $T$-coupons are inserted to connect matching cable
ends. This is because the $T$-coupons can cross arbitrarily
(Figure~\ref{fig:propT}.g) and ``windings'' in the connections are
irrelevant (Figure~\ref{fig:propT}.h).
The invariance of the morphism under the way the projections of the
3-spheres are performed (while respecting the orientation) follows
precisely from the $S^3$ (or $\R^3$ which is the same) isotopy
invariance of the diagrammatics.
For the invariance under the choice of framing we note that a change in
the framing for a given piece of wire induces by construction a
corresponding change in the corresponding piece of wire. These give
rise to additional twists in the projections of this wire piece which
appear as mirror images on both sides of the relevant $T$-coupon. As
the $T$-coupon commutes with any morphism the twists can be pulled
``through'' to the same side of the $T$-coupon where they
``annihilate'' each other, see Figure~\ref{fig:twistcom}.
It remains to show the invariance under the choice of alignment of
ribbons for each cable. 
Any such alignment can be obtained from a given one by inserting an
isotopy in a neighborhood of the 2-sphere (cable).
In the projections this amounts to inserting
a diagram consisting of crossings and twists. This diagram is
inserted on both sides of the $T$-coupon with one being the mirror
image of the other. However, as we can pull any morphism (such as the
inserted diagram) through a $T$-coupon, the two mirror images
annihilate each other, leaving the total diagram invariant.

\subsubsection{Definition of the Partition Function}

We proceed to give the formal definition of LGT for the considered
nonsymmetric settings analogous to Definition~\ref{def:lgt}.

\begin{dfn}
\label{def:qlgt}
Let $\catmod$ be a semisimple pivotal category and $n=2$ or a
semisimple spherical category and $n=3$ or a
semisimple ribbon category and $n=4$. Let $\{\alpha_V\}$ be an
assignment of a complex number to
each isomorphism class of simple object $V$ such that
$\alpha_{V^*}=\alpha_V$. These are called \emph{weights}.
Let $M$ be an oriented cellular manifold of dimension $n$.
This defines a \emph{lattice gauge theory} as follows.

For any choice of orientation and labeling $V_f$ with an equivalence
class of simple objects for each $(n-2)$-cell $f$ we define
$\pf_{V_f}$ to be the value of the projected circuit diagram
constructed above. We call
\[
 \pf\defeq\sum_{V_f} (\prod_f \alpha_{V_f}) \pf_{V_f}
\]
the \emph{partition function}, where the sum runs over all possible
labelings. This does not depend on the chosen orientations of the
$(n-2)$-cells.
\end{dfn}

The independence of $\pf$ on the chosen orientations follows
in the same way as in the symmetric case. This is now due to
Lemma~\ref{lem:pivtr}. In the 2 dimensional case it would also make
sense to induce the orientations of the faces from the
orientation of $M$. Then we can drop the condition
$\alpha_V=\alpha_{V^*}$, making $\pf$ possibly dependent on the
orientation of $M$.

\section{Gauge Symmetry and Gauge Fixing}
\label{sec:gauge}

In this section we consider how the gauge invariance of lattice gauge
theory manifests itself in the diagrammatic formulation. Then
we show how the notion of gauge fixing in conventional lattice gauge
theory translates into the diagrammatic language and generalizes to
the symmetric, pivotal, spherical and ribbon settings. Furthermore, it
turns out to be related to a topological move between cellular
decompositions which is thus an invariance of the partition function.

\subsection{Gauge Symmetry}
\label{sec:gaugesym}

Recall that in conventional lattice gauge theory a gauge
transformation is defined
by assigning a group element $g_v$ to each vertex $v$. This changes a
configuration (i.e.\ an assignment of group elements to edges) as
follows. For a given edge $e$ the assigned group element $h$ is
replaced by $g_{v_1} h g_{v_2}^{-1}$ where $v_1$ and $v_2$ are the
vertices bounding $e$. The order of the vertices is determined by the
orientation of $e$.

Consider the LGT action (\ref{eq:lgtact}).
The effect of a gauge transformation on the function $\sigma$ (and
thus for the characters in the expansion of its exponential)
is precisely an action of the group by conjugation
(\ref{eq:conjact}). Namely, $\sigma$ is conjugated by the group
element assigned by the gauge transformation to the vertex which forms
the starting point for the product over group elements
which is the argument of $\sigma$. Conjugation invariance of $\sigma$
implies gauge invariance.

In the diagrammatic formulation the gauge invariance is much more
implicit. The fact that the diagrams represent intertwiners means that
we have an underlying action of the gauge group ``everywhere''. The
fact that a diagram is closed implies invariance. However, we can
still specifically identify the original gauge invariance.

\begin{figure}
\begin{center}
\input{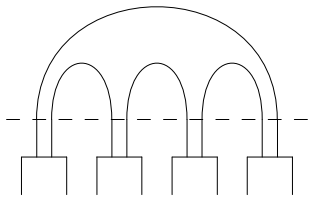}
\end{center}
\caption{
Gauge symmetry at a vertex. Acting with the group at the cut (dashed
line) is equivalent to a gauge transformation.}
\label{fig:gaugesym}
\end{figure}

For
simplicity we consider a gauge transformation that is nontrivial only
at one vertex. Deform the (two dimensionally projected) diagram that
defines the
partition function such that the considered vertex is at the top with
all attached cables leading downward to the remaining diagram, see
Figure~\ref{fig:gaugesym}.
Then we introduce a horizontal cut in the diagram, just on top of the
cables (dashed line). As the diagram above the cut is closed to the
top it is
invariant. Thus, we can act with a group element $g$ on the tensor
product of representations that is represented by the wires crossing
the dashed line without changing the value of the diagram. This action
is simply an action with $g$ on each tensor factor. As the wire pieces
represent functions on the group obtained from the expansion of the
characters, the action with $g$ corresponds to an insertion of $g$
into the evaluation of the function. Each piece of wire crosses the dotted
line twice corresponding to an insertion of $g$ for each of the two
edges connected by the piece of wire. However, the orientation in both
cases is opposite (including the arrow, it points upwards at one
crossing and downwards at the other). Thus, it corresponds to inserting
$g$ on one side and $g^{-1}$ on the other. We recover an ordinary gauge
transformation.

So far we have only talked about the case of ordinary group
symmetries as only in that case gauge transformations can be defined
in the conventional way.
However, the statement of gauge symmetry can more generally be
considered to lie in the fact that we have a diagrammatic formulation
of the partition function. As a diagram determines a morphism in the
relevant category it is covariant by construction. This covariance is
with respect to a group, a supergroup or a quantum group (Hopf
algebra) if the category arises as the category of representations of
the respective object.

\subsection{Gauge Fixing}

As we know from conventional LGT we can use its gauge symmetry to
remove some of
the group integrals in the partition function (\ref{eq:lgtpf}).
The corresponding group variables can be set to the unit element. We
are allowed to do this for the group variables of as many
edges as we like, as long as these do not form any closed loop
\cite{Cre:gaugelat}.

How is this ``gauge fixing'' expressed diagrammatically?
Looking back at expression (\ref{eq:intf})
we see that
removing the integral
means applying the evaluation without the $T$-projector.
Diagrammatically, the projector diagram is simply
removed, i.e.\ replaced by the identity.
In the circuit diagram this means that we can remove
cables by exposing the wires without changing the value of the
diagram. Conventional LGT tells us that
we are allowed to do this as long as the edges for which the
attached cables have been removed do not form any closed loop.

As the gauge invariance is contained in the
diagrammatic formulation we should be able to derive the gauge fixing 
directly diagrammatically -- without recurrence to conventional LGT.
This is indeed the case, as we will show by exploiting the properties
of the $T$-projector (Proposition~\ref{prop:proj} and
Figure~\ref{fig:propT}). This generalizes gauge fixing of conventional
LGT from the group context to the general symmetric setting
as well as to the pivotal, spherical and ribbon settings in the relevant
dimensions.
We start by deriving the identity that enables the gauge fixing on
the purely diagrammatic level. In the ribbon case we assume blackboard
framing.

\begin{figure}
\begin{center}
\input{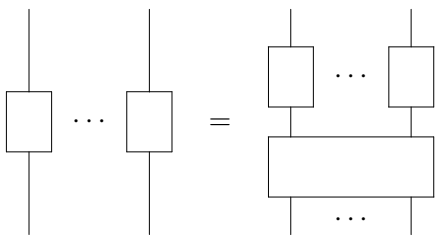}
\end{center}
\caption{Multiple composition property of $T$.}
\label{fig:projmult}
\end{figure}

Let us first observe the additional property of
$T$ depicted in Figure~\ref{fig:projmult}: A tensor product of
$T$-projectors is equal to this same tensor product composed with an 
overall $T$-projector. This follows straightforwardly by multiple
application of properties (c) and (e) of Figure~\ref{fig:propT}.

\begin{figure}
\begin{center}
\input{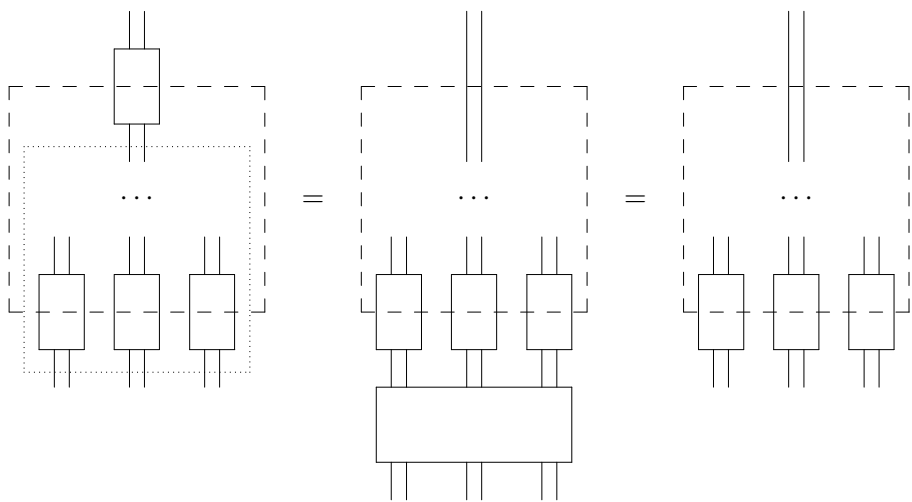}
\end{center}
\caption{Diagrammatic gauge fixing identity.}
\label{fig:gaugefix}
\end{figure}

Consider now a diagram with $T$-coupons (which might arise as the
projection of a circuit diagram). Draw a
closed loop that only intersects $T$-coupons. By
moving the $T$-coupons around we arrive at a diagram as shown in
Figure~\ref{fig:gaugefix} on the left hand side (for the case of four
cables with two wires each). The loop is represented by the dashed
line. The part of the
diagram between the $T$-coupons lies inside the dashed box and is
indicated by three dots. The rest of the
diagram is attached at the top and bottom and lies outside the dashed
box.

Now consider the dotted box.
Whatever lies inside defines a morphism by construction.
Thus, we can apply property (d) of Proposition~\ref{prop:proj} (see
Figure~\ref{fig:propT}) to
the $T$-coupon at the top and this morphism. That is to say we can
exchange the two. The result is shown in the diagram in the middle of
Figure~\ref{fig:gaugefix}. Now, the arrangement of the $T$-coupons at
the bottom
resembles the right hand side of Figure~\ref{fig:projmult}. That is, we
can apply the identity of this figure to arrive at the right hand side
diagram of Figure~\ref{fig:gaugefix}. The result of the operation is
simply the disappearance of the $T$-coupon that was originally at the top.

\subsection{$n$-Cell Fusion Invariance}

It turns out that gauge fixing is much more than an identity that
helps us to simplify diagrams. Assume we are given a diagram that
arises as the projection of a circuit diagram for a cellular
manifold. Assume further that we obtain a new diagram from the
given one by applying a gauge fixing identity (as in
Figure~\ref{fig:gaugefix}).
Remarkably, it turns out that the new diagram is still a
projection of the circuit diagram for the same manifold -- but with
different cellular decomposition. The process that transforms one
cellular decomposition into the other is given by the following
definition.

\begin{dfn}
\label{def:cellfuse}
Let $M$ be a manifold of dimension $n$ with cellular decomposition
$\mathcal{K}$. Let $\mu$ be an $(n-1)$-cell in $\mathcal{K}$ which
bounds two distinct $n$-cells $\sigma$, $\sigma'$. Removing $\mu$,
$\sigma$ and $\sigma'$ from $\mathcal{K}$ while adding the new
$n$-cell $\sigma''\defeq\sigma\cup\mu\cup\sigma'$ leads to a new
cellular decomposition $\mathcal{K}'$ of $M$. We call this process the
\emph{$n$-cell fusion move}.
\end{dfn}

As a consequence the partition function remains invariant under
this move.
To proof our claim let us first consider what the $n$-cell fusion move
means for the circuit diagram. In the context of
Definition~\ref{def:cellfuse}, the
$(n-1)$-cell $\mu$ corresponds to an edge of the associated lattice
and thus a cable of the circuit diagram which is
removed. The wires remain exactly the same however as they correspond
to faces and thus $(n-2)$-cells which are not changed.
That is, the cellular decompositions $\mathcal{K}$ and
$\mathcal{K}'$ have identical circuit diagrams except that in the one for
$\mathcal{K}'$ a cable is removed, exposing the wires.

We need to verify that the projections of the circuit diagram are related
by a gauge fixing identity as depicted in Figure~\ref{fig:gaugefix}.
First, identify the projection of the circuit diagram for
$\mathcal{K}$ with the left hand side diagram of
Figure~\ref{fig:gaugefix} as follows.
$\sigma$ is projected to the interior of the dashed box, $\mu$ to
the top part of the dashed line and $\sigma'$ above it. The dashed
line intersects only cables as it is the projection of the boundary of
$\sigma$. Thus, the diagrammatic identity can be applied which
corresponds to removing the cable that ``pierces'' $\mu$, as
required.

\begin{figure}
\begin{center}
\includegraphics{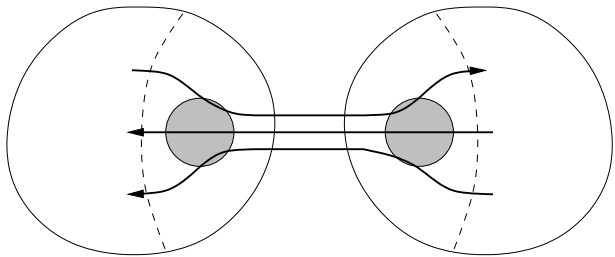}
\end{center}
\caption{Applying the gauge fixing to the projection, removing a
$T$-coupon (cable).}
\label{fig:fixproj}
\end{figure}

\begin{figure}
\begin{center}
\includegraphics{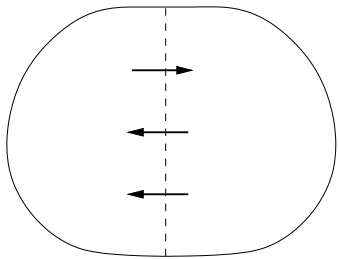}
\end{center}
\caption{Pulling the projected diagram together, so that a projection
of the fused $n$-cell results.}
\label{fig:pfused}
\end{figure}

What remains to check is that the diagram obtained by first
projecting the circuit diagram and then applying the diagrammatic identity
is equivalent to the diagram obtained by first applying the move
(changing the circuit diagram) and then projecting. This is obvious if any
kind of projection (preserving the combinatorics) is allowed, as for
symmetric categories. It is equally obvious in the 2 dimensional
pivotal case. It is less obvious though for the spherical (3d) and
ribbon (4d) cases. As in those cases $n$-cells (more precisely: their
boundaries) are projected separately it is 
sufficient to consider just the projections of (the boundaries of)
$\sigma$, $\sigma'$ and $\sigma''$. We make use of the fact that
the boundaries of $\sigma$ and $\sigma'$ share the $(n-1)$-cell
$\mu$. Thus, we choose the projections of the boundaries of $\sigma$
and $\sigma'$ such that the projections of $\mu$ are identical as
mirror images (as the orientation is reversed). Furthermore, we make
sure that the cable ending in $\mu$ lies near the boundary of each
projection, i.e.\ there is no wire between the cable ending and the
boundary of the projection. (In the spherical 3d-case this is achieved
by choosing the point piercing the boundary 2-spheres close to this
cable. In the ribbon 4d-case this is simply achieved by an appropriate
3-dimensional isotopy.) Also
we make sure in the 4d-ribbon case that there are no crossings inside the
projection of $\mu$, and that blackboard framing applies there (again
by isotopy). The projection we obtain is illustrated by
Figure~\ref{fig:2cellproj}, and after inserting the $T$-coupon by
Figure~\ref{fig:pconnect}. It is now clear that removing the $T$-coupon from
the lines that connect the two projected cells
(Figure~\ref{fig:fixproj}) we can pull the
projections together (until the dashed lines coincide) to obtain precisely
a projection of (the boundary of) $\sigma''$, see
Figure~\ref{fig:pfused}. This completes the proof.

\begin{thm}
Let $M$ be a manifold of dimension $n$ and
$\mathcal{K}$, $\mathcal{K}'$ cellular decompositions of $M$ which are
related by a sequence of $n$-cell fusion moves.
Then, the value of the circuit diagram in a symmetric (or ribbon if
$n=4$, or spherical if $n=3$, or pivotal if $n=2$) category for a
given labeling
and choice of orientation of $(n-2)$-cells is the
same for $\mathcal{K}$ and $\mathcal{K}'$. In particular, the
partition function for both is the same.
\end{thm}

We turn now to the question what the analog of the ``no-loop''
condition for the gauge fixed edges of conventional LGT is. The only
situation that prevents us from removing a cable and fusing the two
$n$-cells that it connects is when these $n$-cells are actually
identical. In that case the dual edge corresponding to the cable forms
indeed a closed loop and we are not allowed to gauge fix
it. Conversely, given a set of dual edges that forms a loop, we cannot
remove all of the cables belonging to it. This is because just before
removing the last cable, the path formed by the gauge fixed edges
would lie completely inside one $n$-cell. Thus, the remaining cable
would belong to an $(n-1)$-cell which bounds this one $n$-cell on both
sides and therefore cannot be removed by $n$-cell fusion.

In the symmetric category case we do not need to think of the lattice
as arising from a cellular complex. In that case we can express the
gauge fixing move directly in terms of the lattice. It corresponds to
removing one edge by identifying all its points so that its two
bounding vertices become one. This is only allowed if these two
bounding vertices are distinct.

\section{Observables}

\label{sec:obs}

\begin{figure}
\begin{center}
\includegraphics{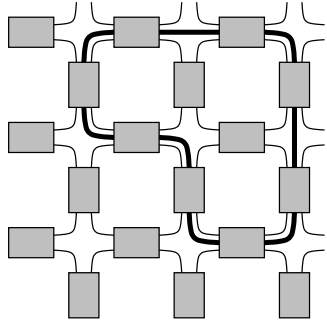}
\end{center}
\caption{A circuit diagram with a Wilson loop (thick line).}
\label{fig:loopnet}
\end{figure}

The standard observables of conventional lattice gauge theory are
Wilson loops or,
more generally, (embedded) spin networks. A Wilson loop $L$ is a
subset of edges of
the lattice that form a closed loop, carry a consistent orientation,
and it has attached the label of an irreducible representation. The
partition function (\ref{eq:lgtpf}) with a Wilson loop inserted takes
the form
\begin{equation}
 \pf[L]=\int (\prod_e \xd g_e)\, \chi_L(\prod_{e\in L} g_e)\, e^{-S},
\label{eq:lgtloop}
\end{equation}
where $\chi_L$ is the character of the irreducible representation
carried by $L$.
The expectation value corresponding to the Wilson loop is then
$\pf[L]/\pf$.
After expanding characters as in Section~\ref{sec:lgt} we arrive at an
expression
\begin{equation}
 \pf[L]=\sum_{V_f} (\prod_f \alpha_{V_f}) \pf_{V_f}[L],
\end{equation}
analogous to (\ref{eq:partexp}).
The summand $\pf_{V_f}[L]$
takes the modified form
\begin{align}
 \pf_{V_f}[L] =
 \int (\prod_e \xd g_e)\, \chi_L(\prod_{e\in L} g_e) \prod_f
 \chi_{V_f}(\prod_{e\in f} g_e) .
\label{eq:ffuncloop}
\end{align}
We proceed as in Section~\ref{sec:lgt} to obtain a diagrammatic
representation. The characters are represented by
diagrams (Figure~\ref{fig:copchar}) and inserted into the lattice as
wires. We only have an extra character $\chi_L$ now, which is inserted
into the lattice as a wire along the edges designated by the data of
the Wilson loop. Then the cables are
inserted on the edges to represent the integrations. The only
difference to Section~\ref{sec:lgt} is that the cables for the edges
of the Wilson loop now include the Wilson loop wire as well.
We obtain a modified circuit diagram, see Figure~\ref{fig:loopnet}.
The construction generalizes to several Wilson loops as well as to
arbitrary spin networks inserted into the lattice along edges (with
intertwiners positioned on vertices).

The generalization from conventional LGT to arbitrary symmetric
categories is immediate by taking the modified circuit diagram as a
definition. The generalization to nonsymmetric categories is less
straightforward. The categorial
structures needed to define the partition function in dimension $n$
are related to isotopy in dimension $n-1$ since it is possible to
confine the wires to the boundaries of the $n$-cells. However, as
there is in general no canonical way of putting a
Wilson loop on the boundaries of the $n$-cells, the categorial
structures needed in the presence of a Wilson loop are those
related to isotopy in dimension $n$. Indeed, we can define the value
of a circuit diagram with Wilson loop in dimension 3 only for ribbon
categories, and with the Wilson loop being framed.
In dimension 4 there seems to be no obvious
definition beyond the symmetric case.

In dimension 2 we can continue to use pivotal categories and the
definition of the partition function extends to include Wilson loop
and spin network observables in the obvious way.

To define the value of the circuit diagram with Wilson loop in
dimension 3,
we modify the construction of Section~\ref{sec:sphlgt} as follows.
As there we put the wire pieces onto the boundaries of the 3-cells. In
contrast, we leave the ribbon that defines the Wilson loop completely
inside the 3-cells, except of course were it pierces a 2-disk that
defines a cable. These intersections of the ribbon with the 2-disks we
choose such that they lie on the boundary of the 2-disks and the
ribbon is aligned with the boundary, facing the outside direction.
While in Section~\ref{sec:sphlgt} we project just
the boundary 2-spheres of the 3-cells onto the plane, now we project
the whole 3-cells. However, we do this in such a way that the
restriction of the projection to the boundary 2-spheres is precisely a
projection of the boundary 2-sphere as in Section~\ref{sec:sphlgt}.
Note that this involves making the same choices as there: a point in
each 2-sphere and a point in each circle bounding a 2-disk (cable).
Thus, after including the $T$-coupons,
we arrive at a diagram that is
exactly the same as the projected circuit diagram obtained in
Section~\ref{sec:sphlgt}, except that we have
extra ribbon pieces in it. Furthermore, these ribbon pieces can have
crossings with the wire pieces. To obtain a proper ribbon diagram we
only have to introduce the blackboard framing for the wire pieces.
The value of the diagram defines the partition function with Wilson
loop. Note that we can think of the framings of the wires as
arising directly from the cellular decomposition. Indeed, frame the
wires of the circuit diagram in the plane of the boundary 2-cells, face
facing outwards from the 3-cell to which they belong. In that case we
are even free to project the 3-cells
without the projections restricting to projections of 2-spheres on the
boundaries.

The proof of the independence of the value of the diagram from the
choices made in obtaining it is almost the same as for the spherical
case without Wilson loops. The main difference is that we now use
the 3-dimensional isotopy properties of ribbon graphs instead of the
2-dimensional isotopy properties in the spherical case. Furthermore,
there is one extra choice we have made, namely where to insert the
Wilson loop ribbon into the boundaries of the 2-disks defining the
cables. However, it is easy to see that a different choice for a given
cable just leads to extra braidings on both sides of the corresponding
$T$-coupon in the final diagram, which are inverse (being mirror
images). As the braiding
is a morphism, it commutes with the $T$-projector (by property (d) in
Figure~\ref{fig:propT}) and
can thus be ``pulled through'' the $T$-coupon and ``annihilated'' with
its inverse braiding. (Compare the proof for the 4-d ribbon case,
Section~\ref{sec:riblgt}.)

This construction
generalizes in the obvious way to ribbon spin networks by
including coupons. Thus, the observables in the 3-dimensional
ribbon case are framed Wilson loops or, more generally, ribbon
spin networks embedded into the manifold.

The types of admissible category for LGT with and without
observables are summarized in Table~\ref{tab:admcat}.

\begin{table}
\begin{center}
\begin{tabular}{|r|l|l|}
\hline
dim. & LGT & LGT + obs. \\
\hline
$\ge 5$ & symmetric & symmetric \\
4 & ribbon & symmetric \\
3 & spherical & ribbon \\
2 & pivotal & pivotal \\
\hline
\end{tabular}
\caption{Admissible types of categories for generalized LGT in
different dimensions.}
\label{tab:admcat}
\end{center}
\end{table}
\section{Boundaries and TQFT}
\label{sec:boundary}

Here we consider how our diagrammatic definition of the partition
function extends to manifolds with boundary.

\begin{figure}
\begin{center}
\includegraphics{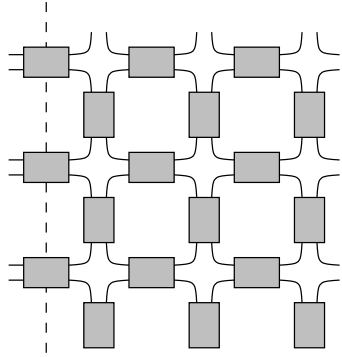}
\end{center}
\caption{The circuit diagram for a cellular manifold with boundary.
The boundary is indicated by the dashed line.}
\label{fig:cwboundary}
\end{figure}

\begin{figure}
\begin{center}
\includegraphics{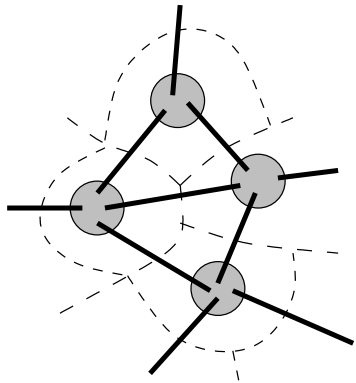}
\end{center}
\caption{The circuit diagram on the boundary. The $(n-1)$-cells are
indicated by dashed lines. They are pierced by cables (represented as
gray disks). The wires in the cables lie at the endpoints of the thick
lines. These lines connect corresponding wire pieces. They are
the edges of the spin network on the boundary.}
\label{fig:boundarysn}
\end{figure}

Let $M$ be a cellular manifold of dimension $n$ (oriented if the
category is nonsymmetric) with boundary
$\partial M$. (That is, a manifold with boundary having a cellular
decomposition of the boundary that extends to a cellular decomposition
of the manifold.)
We can straightforwardly perform the construction of the circuit
diagram in $M$, the only difference to the case without boundary
being that we
now have wire pieces with free ends on the boundary. These wire
ends are actually inside cables that end on the boundary. See
Figures~\ref{fig:cwboundary} and \ref{fig:boundarysn} for
illustration. After projection, done in the usual way we now obtain a
morphism not from
the unit object $\one$ to itself, but between the unit object and the
object associated with the boundary. We have the choice whether we want
to consider this boundary object to determine the domain or image of
the morphism. Diagrammatically, this is just the choice of writing the
diagram such that all ``loose'' cable ends are aligned on the top
(domain) or on the bottom (image). Note that changing this choice
exchanges objects and dual objects as the arrows on the wires change
their direction with respect to the vertical.

Assume for the moment that we have chosen the boundary object to lie
in the image of the morphism. What is this boundary object? We have
one cable for each vertex
$v$ of (the lattice associated with) the boundary (or $(n-1)$-cell).
Each
cable has a bunch of wires, with each piece of wire corresponding to one
edge $e$ meeting in $v$ ($(n-2)$-cell bounding the
$(n-1)$-cell ``before identification''). Thus we have a tensor product
$\bigotimes_{v\in\partial e} V_e$ as the object defined by the wires
in that cable. The object corresponding to the whole boundary
$\partial M$ is the tensor product of all these objects. Thus, the
(open) circuit diagram gives rise to a morphism from $\one$ to this tensor
product. However, we have not used the fact that there are cables
around the wires on the boundary. Due to the decomposition property of
the cables (Proposition~\ref{prop:proj}),
the total morphism decomposes into a tensor product of morphisms
\begin{equation}
I_v:\one\to \bigotimes_{v\in\partial e} V_e
\label{eq:bmorph}
\end{equation}
for each cable (and dual vertex $v$). (Note that this defines
the individual morphisms $I_v$ only up to scale. On the other hand the
order of the $I_v$ in the tensor product does not matter due to the
identity $\one\tens V=V\tens\one$ for the category.)
If we have made the other choice, namely that the boundary object
should lie in the domain of the morphism we would have obtained a
tensor product of morphisms of the type
\begin{equation}
I'_v:\bigotimes_{v\in\partial e} V_e\to\one .
\label{eq:bmorphd}
\end{equation}

We can also think of the morphisms (\ref{eq:bmorph}) or
(\ref{eq:bmorphd}) as states on the boundary, as they form vector
spaces and can be paired in the obvious way.
More precisely, a state would also include the specification of the
labelings of the edges. Thus, the complete description of a state
would be a labeling of the edges with arrows and simple objects, and a
labeling of the vertices by morphisms (as specified by
(\ref{eq:bmorph}) or (\ref{eq:bmorphd})) between the objects that
label the incident edges. In fact, this is just an embedded
spin network. We recover the well known picture of spin
networks as states on the boundary of spin foams, see
\cite{Bae:spinfoams}.
Note that to conform to our definition in Section~\ref{sec:diag} in the
4 dimensional ribbon case we would expect the spin network to be a
ribbon graph. We can indeed think of it in this way. To this end
consider the circuit diagram embedded into the 3 dimensional
subcomplex as described in Section~\ref{sec:qlgt}. The ribbons of the
spin network are then obtained from the wire pieces that touch the
boundary by removing their parts inside the manifold and gluing them
together at the 2-cells on the boundary.

The states form a vector space since the morphisms form vector spaces
and we can take the direct sum over the labelings of edges by
objects. Thus, for a cellular manifold $N$ of dimension $n-1$
we define the state space to be
\begin{equation}
 \cH_N\defeq
 \bigoplus_{V_e} \left(\bigotimes_{v} 
 \Mor\left(\bigotimes_{v\in\partial e} V_e, \one\right)\right) .
\label{eq:statesp}
\end{equation}
The dual state space $\cH^*_N$ is defined in the obvious
way, by exchanging the 
arguments in $\Mor$. (Note that this also exchanges objects
with dual objects as diagrams are turned upside down.)
The pairing between a state and a dual state is the obvious one if the
labelings of edges coincide. Otherwise the pairing is defined to be zero.

Summing over all labelings, a cellular manifold $M$ with boundary
$\partial M$ gives
rise to a state (or dual state) and thus to a linear map
$\C\to\cH_{\partial M}$. (Note that in the case of infinitely many
inequivalent simple objects we need to consider a completion of the
direct sum in (\ref{eq:statesp}) to make this map algebraically well
defined.) We
use the usual weights $\alpha_V$
inside the manifold while we use a square root $\sqrt{\alpha_V}$ on
the boundary, i.e., for the faces piercing the boundary.
Dually, we can also think of this as giving rise to a linear form
$\cH_{\partial M}\to \C\cong\mathrm{Mor}(\one,\one)$, by composition
(pairing) with a spin network state.
If $\partial M$ consists of several connected components we can make
different choices for the different components as to correspond to domain
or image of such linear maps. In particular, let us assume that
$\partial M$ consists of
two components $\partial M=\partial M_1\cup \partial M_2$.
Now defining $\partial M_1$ to correspond to the domain and $\partial
M_2$ to the image we obtain a linear map
\begin{equation}
 \Omega_{\partial M}:\cH_{\partial M_1}\to\cH_{\partial M_2}
\end{equation}
in the obvious way.

Now assume we have two cellular manifolds $M$, $M'$ with boundaries
$\partial M=\partial M_1\cup\partial M_2$ and
$\partial M'=\partial M'_1\cup\partial M'_2$ respectively, as well as
a cellular homeomorphism identifying $\partial M_2\cong \partial
M'_1$. Gluing the manifolds
together along $\partial M_2$, $\partial M'_1$ we obtain a new one
$M''=M\cup M'$ with
boundary $\partial M''=\partial M_1\cup\partial M'_2$.
This gives rise to linear maps $\Omega_{M}$, $\Omega_{M'}$ and
$\Omega_{M''}$ satisfying the composition property
\begin{equation}
 \Omega_{M''}=\Omega_{M'}\circ\Omega_{M} .
\label{eq:cobcomp}
\end{equation}
This is because the circuit diagram for $M''$ is just the same as the ones
for $M'$ and $M$ attached to each other. This is also true for the
projections (in the symmetric as well as the nonsymmetric cases)
which define its values, as the attachment is only between
cables/$T$-coupons.
Note that we use the projector property (c) of the
$T$-coupon (Figure~\ref{fig:propT}) as in attaching the circuit
diagrams we have cables on both sides which we then ``glue'' to single
cables.
Furthermore, the weights $\sqrt{\alpha_V}$ on the boundary recombine
to the usual weights $\alpha_V$.

The state space of spin networks is complete in the following
sense. Take the situation above with the manifolds $M$, $M'$ etc. The
spaces $\cH_{\partial M_2}$ and $\cH_{\partial M'_1}$ are
identified. Taking a basis $|\psi\rangle$ of $\cH_{\partial M'_1}$
(consisting of a basis of 
morphisms for each labeling of edges) and the dual basis
$\langle\psi|$ of $\cH^*_{\partial M_2}$ we can write this as
\begin{equation}
 \sum_\psi\Omega_{M'}|\psi\rangle\, \langle\psi|\Omega_{M}
 =\Omega_{M'}\circ\Omega_{M} .
\end{equation}
Diagrammatically, this is just the insertion of decompositions
for the cables crossing the common boundary, as
well as writing the sum over labelings of the wires crossing the
boundary explicitly.

What we arrive at is ``almost'' a topological quantum field
theory (TQFT). The topological objects are cellular manifolds of
dimension $n-1$ and their cellular cobordisms. This forms ``almost'' a
category. (The notion of identity 
morphism is lacking.) On the other hand we have the category of vector
spaces and linear maps. What we obtain is thus ``almost'' a functor
from the ``cellular cobordism category'' to the category of vector
spaces by assigning state spaces to cellular manifolds and linear maps to
cellular cobordisms. In particular, the crucial composition property
(\ref{eq:cobcomp}) is satisfied. In the topologically invariant case
(i.e., when the partition function is independent of the cellular
decomposition, see Section~\ref{sec:topinv}) one can forget about the
cellular decomposition of the cobordism. The identity in the cobordism
category
is then the ``cylindrical'' cobordism; for a manifold $N$, this is
$I\times N$ with $I$ a closed interval. Quotienting the state space
$\mathcal{H}_N$ by the kernel of $\Omega_{I\times N}$ then gives rise
to a TQFT. See \cite{TuVi:inv3} for this kind of quotient construction.

It is straightforward to combine boundaries
and Wilson loop (or spin network) observables. Now the manifolds are
allowed to have Wilson loops embedded in them, which may end
on the boundary. Thus, the boundaries are cellular manifolds with
possible extra labels on their vertices indicating the object
label of a Wilson loop $L$ piercing the boundary at this vertex. The
morphisms (and states) on the boundary are modified so as to include the 
extra objects. The change is just the inclusion of the the object as
an extra factor $V_L$ in the tensor product $(\bigotimes_{v\in\partial
e} V_e)\tens V_L$ for this vertex. Everything else in the above
construction works as before. Thus, we obtain ``almost'' a TQFT. Now
it is defined on the ``almost category'' of cellular manifolds with
possible extra object labels on the dual vertices. The cobordisms are
now cellular manifolds which have Wilson loops (or more generally spin
networks) embedded, as discussed in Section~\ref{sec:obs}, with
corresponding labels. This works now for symmetric categories in any
dimensions, and for ribbon categories in dimension 3. In fact, in the
latter case, as the Wilson loops are ribbons, there is slightly more
structure not only on the cobordisms (as already discussed in
Section~\ref{sec:obs}), but also on the boundaries.
The ends of the Wilson loops on the 
boundary need to be considered as carrying a direction which
determines the framing.
This is similar to a situation considered in \cite{Tur:qinv},
where a TQFT with Wilson loops in dimension 3 is constructed from the
surgery invariants of \cite{ReTu:inv3qg}.

Finally, (without working out any details) we mention that given extra
complex structure on the
category, we can make the state spaces into Hilbert spaces. For
example, if the category is the category of representations of a
compact Lie group we can identify dual representations as conjugate
representations. Thus, we obtain Hilbert space structures on the
representations and in turn on the relevant intertwiner spaces
(between the trivial and an arbitrary representation).
\section{Special Cases}

\label{sec:specas}

\subsection{Spin Foams and $nj$-Symbols}
\label{sec:spinfoams}

\begin{figure}
\begin{center}
\input{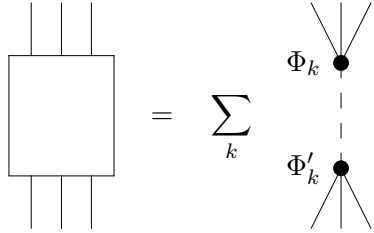}
\end{center}
\caption{Decomposition of the $T$-morphism on a tensor product.}
\label{fig:njdec}
\end{figure}

We discuss here how the conventional picture of spin foam
models employing polyhedral ``recoupling diagrams'' is
recovered. Consider a cable
around a number of wires carrying object labels $V_1,\dots,V_n$
representing the morphism $T_{V_1\tens\cdots\tens V_n}$.
By definition (Proposition~\ref{prop:proj}),
we can decompose it into morphisms
$\Phi_i:V_1\tens\cdots\tens V_n\to\one$ and
$\Phi_i':\one\to V_1\tens\cdots\tens V_n$
such that
\begin{equation}
 T_{V_1\tens\cdots\tens V_n}=\sum_k \Phi_k' \Phi_k .
\label{eq:decable}
\end{equation}
We depict this diagrammatically as
in in Figure~\ref{fig:njdec}. The morphisms are here represented by coupons
which are shrunk to dots. Note that the dashed line representing the
unit object $\one$ is normally omitted and here only drawn for
illustration.

We can introduce such a decomposition for every cable in a
circuit diagram. The resulting graph then consists of disconnected
polyhedral diagrams,
one for each $n$-cell (or vertex of the associated lattice). The lines of
the polyhedra are
the wires while its corners arise where the wires entered a cable. The
partition function now has an extra summation besides the one over
labelings of faces with simple objects. This is
the summation over the decomposition (\ref{eq:decable})
for every edge. Thus, we can express the partition function as
\begin{equation}
 \pf=\sum_{V_f} (\prod_f \alpha_{V_f})
\sum_{\Phi_e} \prod_v A_v (V_f, \Phi_e) .
\label{eq:spinfoam}
\end{equation}
Here, $\Phi_e$ denotes a morphism between the tensor product of
objects corresponding to the wires on the edge $e$, and we sum
over labelings with such morphisms as prescribed by
(\ref{eq:decable}). $A_v$ denotes the value of the polyhedral diagram
arising at the vertex $v$. It depends on the labelings of
the faces and edges that meet in $v$.

Formula (\ref{eq:spinfoam}) is essentially the general definition of a
spin foam model, except that one usually requires the weight
$\alpha_V$ to be the dimension of the representation $V$, i.e.,
$\alpha_{V}\defeq\dim_{V}$ (or for general categories
$\alpha_{V}\defeq\cdim_{V}$).
For general spin foam models, one has some freedom in
defining the ``vertex amplitude'' $A_v$. The one we obtain here
(defined by the polyhedral diagrams as described above and with the
special choice of weights) defines the
spin foam models of BF-type. These are topological and give rise to
state sum invariants, see the next section.

Usually one chooses bases of the decompositions (\ref{eq:decable}) in
a globally coherent
way for the whole category so that they can be indexed.
However, there is no canonical way of
doing this. One can compare this to a choice of
``coordinates''. In this sense, the circuit diagram formulation of the
partition function is ``coordinate-free''. In contrast, for a
definition that starts out from (\ref{eq:spinfoam}), one would have to show
independence of $\pf$ under the choice of bases.

The spin foam approach (and related state sum models) are normally
restricted to a simplicial decomposition of the manifold. This has the
effect that the number of edges meeting in a vertex and the number
of faces bounded by an edge is a fixed number just depending
on the dimension of the manifold. This means that just one type of
polyhedral diagram with fixed number of edges and vertices
appears. Thus, one has only one type of ``recoupling symbol'' as
such polyhedral diagrams are called. For example, in
dimension 3 this is a $6j$-symbol and in dimension 4 a $15j$-symbol.
The standard approach at showing that a state sum of the type
(\ref{eq:spinfoam}) is well defined (or even a topological invariant)
is by using properties of these recoupling symbols.
Of course, this seems rather hopeless if infinitely many types of
recoupling symbols can occur, hence the restriction to simplicial
decompositions.

\subsection{Weak Coupling Limit, State Sum Invariants and BF-Theory}
\label{sec:topinv}

For conventional LGT (with a compact Lie group $G$) one requires the
local action $\sigma$
(\ref{eq:lgtact}) to recover the continuum action of Yang-Mills theory
in the limit of small lattice spacing. In particular, $\sigma$ will be
a function of the coupling constant $\lambda$ of the continuum
theory. Then, the Boltzmann weight $e^{-\sigma(\lambda,g)}$ tends to
the delta function $\delta(g)$ in the weak coupling limit $\lambda\to
0$.
Thinking of the partition function as a path integral over
connections, this means that only flat connections contribute.
In terms of the weights, this limiting case is $\alpha_V=\dim V$.
In our formulation, there is nothing manifestly singular about this
case. However, the partition function $\pf$ will in general not
converge anymore. An alternative way to obtain this partition function
is as a discretization of BF-theory. Integrating out the $B$-field
yields the delta function on the connection.

The attractive feature about BF-theory is that it is a topological
theory. That means, the discretized partition function depends
only on
the topology of the manifold.
It is the same (up to a factor) for any cellular
decomposition of it. This extends beyond the group case
and for the various categories we have considered the ``topological''
weight is given by $\alpha_V=\cdim V$.
Indeed, our construction recovers the various state sum invariants in
this case.
While for group representations
$\cdim V= \dim V$, we have $\cdim V=\sdim V$ (the super-dimension) in
the supergroup case and $\cdim V =\qdim V$ (the quantum dimension) in
the quantum group case.

In 2 dimensions the situation is rather simple (as LGT is then
solvable). Consider a compact connected oriented 2-manifold $M$ with
some cellular decomposition and a pivotal category $\catmod$.
Induce the orientations of the faces 
of the associated lattice from $M$.
Project onto the plane with positive (anti-clockwise) orientation and
set $\alpha_V\defeq\cdim_- V$. We apply the
identity of Proposition~\ref{prop:dualid} to all cables of the circuit
diagram. If any two adjacent faces
are labeled with different objects, the partition function
vanishes. Thus, we have only one sum over simple objects. All the
cables are replaced with the diagram at the right hand side of
Figure~\ref{fig:dualid}. Thus, we just obtain a bunch of loop
diagrams with negative (clockwise) oriented arrows. Indeed, we obtain one
loop for each vertex (2-cell), one
inverse loop for each edge (1-cell) and one loop for each face
(0-cell), from the weight. Consequently,
\begin{equation}
 \pf=\sum_V (\cdim_- V)^\chi
\end{equation}
with $\chi=n_2-n_1+n_0$ the Euler characteristic of $M$ ($n_i$ the
number of $i$-cells). (Note that we could have equally chosen
$\alpha_V\defeq\cdim_+ V$ and projected with negative orientation. This has
the same effect as interchanging $V$ and $V^*$. Thus, the resulting
$\pf$ is the same.)

In higher dimensions the above mentioned factor has to be taken into
account to obtain a true invariant. The invariant is
\begin{equation}
\pfm\defeq\kappa^{-\chi_L} \pf\quad\text{with}\quad
 \kappa\defeq\sum_V (\cdim V)^2
\end{equation}
and $\chi_L$ is the ``Euler characteristic'' of the associated
lattice. That is $\chi_L\defeq n_v-n_e+n_f=n_d-n_{d-1}+n_{d-2}$ in
dimension $d$. As $\chi_L=\chi$ an invariant in 2 dimensions there was
no need for this factor in that case.

In 3 dimensions for $SU(2)$
we recover the Ponzano-Regge model \cite{PoRe:limracah}, which yields
a divergent partition function as the sum over representation labels
is infinite. For $SU_q(2)$ at a root of unity (giving rise to a
ribbon category, see next section) we recover the
Turaev-Viro state sum \cite{TuVi:inv3}, which defines interesting
3-manifold invariants. The generalization to spherical categories
(to which our definition extends) was achieved by Barrett and Westbury
\cite{BaWe:sphcat}.
The proof of topological invariance in our framework turns out to be
surprisingly simple, as it can be cast purely in the diagrammatic
language \cite{GiOePe:diagtop}. In fact, while previous proofs have
employed simplicial
decompositions, the generalization to cellular decompositions makes
the proof even simpler. This is because it can be cast in terms of
moves between cellular decompositions which correspond to
``elementary'' diagrammatic identities.

In 4 dimensions, the group $SU(2)$ yields Ooguri's analogue of the
Ponzano-Regge model \cite{Oog:toplat}. The quantum group $SU_q(2)$
yields the invariant of 4-manifolds of Crane and Yetter
\cite{CrYe:4dtqft}, later generalized to ribbon categories
\cite{CrKaYe:inv4}. The proof of topological invariance in our
framework should be very similar to the 3 dimensional one.

\subsection{Modular Categories and Chain Mail}

Interesting (finite and non-trivial) examples of state sum invariants
on the one hand
and models of quantum gravity with cosmological constant
\cite{MaSm:qdefqgrav} on
the other hand are obtained
from $q$-deformed groups at roots of unity. These give rise to
quasimodular categories (in the terminology of Turaev \cite{Tur:qinv})
with finitely many equivalence classes of simple objects. Although
these categories are not semisimple, they can be turned into
semisimple ones through a process of ``purification''. This is a
quotient construction on the morphism spaces \cite{Tur:qinv}.
The obtained categories are \emph{modular}, a special case of ribbon
categories with a nondegeneracy condition on the braiding.

\begin{figure}
\begin{center}
\input{figures/fig_modT}
\caption{Identity for the $T$-morphism in a modular
category. The sum runs over equivalence classes $V$ of simple
objects. The arrow directions on the loops are irrelevant and
blackboard framing is implied.}
\label{fig:modT}
\end{center}
\end{figure}

For modular
categories the $T$-morphism can be expressed as a sum over diagrams
with a loop going round a line,
see Figure~\ref{fig:modT}. ($\kappa$ is defined as above.) Consider
the 3 dimensional setting specialized to modular categories.
The circuit diagram can be constructed
as a ribbon diagram freely embedded into the cellular manifold (not
restricted to the 2-dimensional subcomplex), see Section~\ref{sec:obs}.
As the cables of the embedded circuit diagram
are disks (shortened cylinders), we can use the above identity to
replace them by loops going round the wire strands.
This converts the circuit diagram into a pure ribbon link.
We obtain one extra summation over simple objects for each replaced
cable. $\pf_{V_f}$ of Definition~\ref{def:qlgt} is thus decomposed as
\begin{equation}
 \pf_{V_f}=\kappa^{-n_e}\sum_{V_e} \pf_{V_f,V_e},
\end{equation}
with $n_e$ the number of edges and the sum ranging over all labelings
$V_e$ 
of edges with equivalence classes of simple objects. The value
$\pf_{V_f,V_e}$ is now given by the Reshetikhin-Turaev invariant
\cite{ReTu:inv3qg} (in
its TQFT normalization) of
the labelled ribbon link in the given manifold.
This is (in the topological case) essentially a categorial analogue of
Robert's Skein theoretic ``chain mail'' construction of the
Turaev-Viro invariant \cite{Rob:skeintuvi}.

\subsection{Comparison with Previous Generalizations of LGT}

In the 3-dimensional case Boulatov put forward a proposal for
$q$-deformed LGT \cite{Bou:qlgt}. Indeed, this proposal amounts
essentially to
the chain mail construction of modular LGT considered above.

In the 4-dimensional case Pfeiffer recently constructed a ribbon
category generalization of LGT for simplicial
decompositions of the underlying manifold \cite{Pfe:4dlgtrib}.
Our partition function specializes to his in the simplicial case.
However, as general LGT is not topological, the generalization to
cellular decompositions is a substantial improvement. For example,
hypercubic lattices (commonly used in LGT) arise from cellular
decomposition that are not simplicial.

\section{Outlook}
\label{sec:outlook}

In this closing section we discuss several possible developments
suggested by the present work.

For (ordinary) LGT, the diagrammatic formulation of the partition
function introduced here is a further step in
developing the dual model \cite{OePf:dualgauge}.
In the strong coupling regime, contributions to the partition function
with ``small'' representation labels dominate. The diagrammatic
techniques introduced here should help to deal with those
contributions in order to extract a strong-coupling expansion.
On the other hand, the weak coupling regime is equally accessible
through our formalism. Indeed, the proof of topological invariance in
the limit employs elementary diagrammatic identities
\cite{GiOePe:diagtop}. The
``expansion'' of these identities might thus lead to an ``expansion''
of LGT around this topological limit. This would provide a new type of
weak coupling expansion (not destroying the global structure of the
gauge group) possibly shedding new light on the continuum limit.

The generalization of LGT beyond groups might be useful in several
ways. First of all, it makes it possible to put supersymmetric
gauge theories on the lattice. It would be interesting to see how
improvements of convergence and renormalizability of such theories
manifest themselves on the lattice. Going further, LGT with quantum
gauge groups could be something very natural. This is suggested by the
(related) observations that ``quantization'' of the group can occur
both in making sense of a divergent path integral \cite{Wit:qftjones}
or in introducing a term in the Lagrangian which corresponds to a
cosmological constant \cite{MaSm:qdefqgrav}. Indeed, the
regularizing effect of
$q$-deformed groups at roots of unity is immediately apparent in our
LGT framework, as it makes the set of equivalence classes of simple
objects finite and thus the partition function manifestly finite and
well defined.

For BF-theory one immediate application of our formalism is
3 dimensional quantum supergravity. Indeed, in the same way that
BF-theory with gauge group $SO(3)$ or $SO(2,1)$ describes pure
gravity in 3 dimensions, BF-theory with $OSp$ supergroups
describes supergravity in 3 dimensions
\cite{AcTo:cssugra}. Furthermore, in the same way as quantum
BF-theory is also in higher dimensions a starting point for quantum
gravity one could consider quantum BF-theory with the relevant
supergroup a starting point for quantum supergravity. In particular,
this might be of interest in string theory, where 11-dimensional
supergravity is considered one limit of the conjectured M-theory.
A promising model for pure quantum gravity in 4 dimensions was
proposed by Barrett and Crane \cite{BaCr:relsnet}. This is based on a
modification of BF-theory in its spin foam
formulation. Nevertheless, the Barrett-Crane model can still be
expressed in the diagrammatic language introduced here. Thus, the
diagrammatic methods here might help in understanding and developing this
model and its ``relatives''.

An open problem in approaches to quantum gravity is how to perform a
``sum over topologies''. Recently, a proposal has been made to generate
spin foams (i.e., essentially topologies) as Feynman
graphs of a quantum field theory of fields living on the gauge group
\cite{ReRo:stfeyn}. On the other hand it has been shown in
\cite{Oe:bqft} how Feynman diagrams can be rigorously considered as
diagrams denoting morphisms (in the sense of
Section~\ref{sec:diag}) in the category of representations of the
symmetry group of the quantum field theory.
Thus, for such generating field theories we
obtain immediately a representations of the emerging spin foams
(space-times) in terms of the diagrammatics introduced here (using, in
particular Section~\ref{sec:groups}). 
Furthermore, the main emphasis in \cite{Oe:bqft} was the
generalization to braided categories (similar to ribbon
categories). Thus, this provides a way to extend such generating field
theories to quantum groups in the necessary absence of additional
topological input (as we required in Section~\ref{sec:qlgt}).

\subsection*{Acknowledgements}

I would like to thank H.~Pfeiffer, A.~Perez, F.~Girelli and
M.~Reisenberger for valuable discussions and comments on the
manuscript. This work was supported by a NATO fellowship grant.

\bibliographystyle{amsordx}
\bibliography{stdrefs}

\begin{thebibliography}{10}

\bibitem{Sav:duality}
R.~Savit, {\em Duality in field theory and statistical systems}, Rev. Mod.
  Phys. {\bf 52} (1980), 453--487.

\bibitem{Rei:worldsheet}
M.~P. Reisenberger, {\em Worldsheet formulations of gauge theories and
  gravity}, Preprint gr-qc/9412035, 1994.

\bibitem{OePf:dualgauge}
R.~Oeckl and H.~Pfeiffer, {\em The dual of pure non-Abelian lattice gauge
  theory as a spin foam model}, Nucl. Phys. {\bf B 598} (2001), 400--426.

\bibitem{Wit:qftjones}
E.~Witten, {\em Quantum Field Theory and the Jones Polynomial}, Commun. Math.
  Phys. {\bf 121} (1989), 351--399.

\bibitem{Bou:qlgt}
D.~V. Boulatov, {\em $q$-deformed lattice gauge theory and $3$-manifold
  invariants}, Int. J. Mod. Phys. {\bf A 8} (1993), 3139--3162.

\bibitem{Pfe:4dlgtrib}
H.~Pfeiffer, {\em Four-dimensional lattice gauge theory with ribbon categories
  and the Crane-Yetter state sum}, J. Math. Phys. {\bf 42} (2001), 5272--5305.

\bibitem{Bae:spinfoams}
J.~C. Baez, {\em Spin foam models}, Class. Quantum Grav. {\bf 15} (1998),
  1827--1858.

\bibitem{Wit:gravsolv}
E.~Witten, {\em $2+1$-dimensional gravity as an exactly soluble system}, Nucl.
  Phys. {\bf B 311} (1988/89), 46--78.

\bibitem{PoRe:limracah}
G.~Ponzano and T.~Regge, {\em Semiclassical limit of Racah coefficients},
  Spectroscopic and Group Theoretical Methods in Physics (F.~Bloch et~al.,
  ed.), North-Holland, Amsterdam, 1968.

\bibitem{MaSm:qdefqgrav}
S.~Major and L.~Smolin, {\em Quantum deformation of quantum gravity}, Nucl.
  Phys. {\bf B 473} (1996), 267--290.

\bibitem{BaCr:relsnet}
J.~W. Barrett and L.~Crane, {\em Relativistic spin networks and quantum
  gravity}, J. Math. Phys. {\bf 39} (1998), 3296--3302.

\bibitem{Bae:introsfoam}
J.~C. Baez, {\em An Introduction to Spin Foam Models of $BF$ Theory and Quantum
  Gravity}, Geometry and quantum physics (Schladming, 1999), Lecture Notes in
  Phys., no. 543, Springer, Berlin, 2000, pp.~25--93.

\bibitem{ReTu:inv3qg}
N.~Yu. Reshetikhin and V.~G. Turaev, {\em Invariants of 3-manifolds via link
  polynomials and quantum groups}, Invent. Math. {\bf 103} (1991), 547--597.

\bibitem{TuVi:inv3}
V.~G. Turaev and O.~Ya. Viro, {\em State sum invariants of $3$-manifolds and
  quantum $6j$-symbols}, Topology {\bf 31} (1992), 865--902.

\bibitem{BaWe:invplm}
J.~W. Barrett and B.~W. Westbury, {\em Invariants of piecewise-linear
  $3$-manifolds}, Trans. Amer. Math. Soc. {\bf 348} (1996), 3997--4022.

\bibitem{CrYe:4dtqft}
L.~Crane and D.~Yetter, {\em A categorical construction of 4d topological
  quantum field theory}, Quantum Topology (Dayton, 1992), Series on Knots and
  Everything, vol.~3, World Scientific, Singapore, 1993, pp.~120--130.

\bibitem{CrKaYe:inv4}
L.~Crane, L.~H. Kauffman, and D.~N. Yetter, {\em State-Sum Invariants of
  4-Manifolds}, J. Knot Theory Ram. {\bf 6} (1997), 177--234.

\bibitem{FrYe:braidcat}
P.~J. Freyd and D.~N. Yetter, {\em Braided Compact Closed Categories with
  Applications to Low Dimensional Topology}, Adv. Math. {\bf 77} (1989),
  156--182.

\bibitem{ReTu:ribboninv}
N.~Yu. Reshetikhin and V.~G. Turaev, {\em Ribbon graphs and their invariants
  derived from quantum groups}, Commun. Math. Phys. {\bf 127} (1990), 1--26.

\bibitem{BaWe:sphcat}
J.~W. Barrett and B.~W. Westbury, {\em Spherical categories}, Adv. Math. {\bf
  143} (1999), 357--375.

\bibitem{Mac:categories}
S.~{Mac~Lane}, {\em Categories for the Working Mathematician}, 2nd ed.,
  Springer, New York, 1998.

\bibitem{Pen:combst}
R.~Penrose, {\em Angular momentum: an approach to combinatorial space-time},
  Quantum Theory and Beyond (T.~Bastin, ed.), Cambridge University Press,
  Cambridge, 1971, pp.~151--180.

\bibitem{Tur:qinv}
V.~G. Turaev, {\em Quantum Invariants of Knots and 3-Manifolds}, de Gruyter,
  Berlin, 1994.

\bibitem{CaSeMa:liegroups}
R.~Carter, G.~Segal, and I.~MacDonald, {\em Lectures on Lie Groups and Lie
  Algebras}, Cambridge University Press, Cambridge, 1995.

\bibitem{Maj:qgroups}
S.~Majid, {\em Foundations of Quantum Group Theory}, Cambridge University
  Press, Cambridge, 1995.

\bibitem{Oe:qgeosusy}
R.~Oeckl, {\em The Quantum Geometry of Supersymmetry and the Generalized Group
  Extension Problem}, to appear in J. Geom. Phys., Preprint hep-th/0106122,
  2001.

\bibitem{Mas:algtop}
W.~S. Massey, {\em A Basic Course in Algebraic Topology}, Springer, New York,
  1991.

\bibitem{Cre:lgt}
M.~Creutz, {\em Quarks, gluons and lattices}, Cambridge University Press,
  Cambridge, 1983.

\bibitem{Cre:gaugelat}
M.~Creutz, {\em Gauge fixing, the transfer matrix, and confinement on a
  lattice}, Phys. Rev. {\bf D 15} (1977), 1128--1136.

\bibitem{GiOePe:diagtop}
F.~Girelli, R.~Oeckl, and A.~Perez, {\em Spin foam diagrammatics and
  topological invariance}, Class. Quantum Grav. {\bf 19} (2002), 1093--1108.

\bibitem{Oog:toplat}
H.~Ooguri, {\em Topological lattice models in four dimensions}, Mod. Phys.
  Lett. {\bf A 7} (1992), 2799--2810.

\bibitem{Rob:skeintuvi}
J.~Roberts, {\em Skein theory and Turaev-Viro invariants}, Topology {\bf 34}
  (1995), 771--787.

\bibitem{AcTo:cssugra}
A.~Ach\'ucarro and P.~K. Townsend, {\em A Chern-Simons action for
  three-dimensional anti-de~Sitter supergravity theories}, Phys. Lett. {\bf B
  180} (1986), 89--92.

\bibitem{ReRo:stfeyn}
M.~Reisenberger and C.~Rovelli, {\em Spacetime as a Feynman diagram: the
  connection formulation}, Class. Quant. Gravity {\bf 18} (2001), 121--140.

\bibitem{Oe:bqft}
R.~Oeckl, {\em Braided Quantum Field Theory}, Commun. Math. Phys. {\bf 217}
  (2001), 451--473.

\end{thebibliography}
\end{document}